\documentclass[12pt]{article}

\usepackage{amssymb}

\usepackage{epsfig}
\usepackage{graphicx}
\usepackage[hang]{caption}
\usepackage[normalsize,it,hang]{subfigure}
\usepackage{soul,xcolor,enumitem}

\usepackage{amsmath}
\usepackage{url}

\newtheorem{rmk}{Remark}

\begin{document}




\title{Active queue management \\ for alleviating Internet congestion \\ via a nonlinear differential equation \\ with a variable delay}%
\maketitle

\begin{quote} \small
  \textbf{Hugues Mounier}$^1$, %
  \textbf{C\'edric Join}$^{2,5}$, %
  \textbf{Emmanuel Delaleau}$^3$ %
  \textbf{and} %
  \textbf{Michel Fliess}$^{4,5,*}$ \\ %
  \vspace{0.5\baselineskip}%
  
  $^1$ L2S (CNRS, UMR 8506), Université Paris-Saclay, \\
  Centrale Sup\'elec, 3 rue Joliot-Curie, 91192 Gif-sur-Yvette, France \\
  \texttt{hugues.mounier@universite-paris-saclay.fr} \\[.5\baselineskip]
  %
  $^2$ CRAN (CNRS, UMR 7039), Université de Lorraine, \\
  Campus Aiguillettes, B.P. 70239, Vand{\oe}uvre-lès-Nancy, 54506, France \\
  \texttt{cedric.join@univ-lorraine.fr, cedric.join@alien-sas.com} \\[.5\baselineskip]
  $^3$ ENI Brest, UMR CNRS 6027, IRDL, F--29\,200, France \\
  \texttt{emmanuel.delaleau@enib.fr} \\[.5\baselineskip]
  $^4$ LIX (CNRS, UMR 7161), École polytechnique, 91128 Palaiseau, France \\
  \texttt{michel.fliess@polytechnique.edu, michel.fliess@swissknife.tech, michel.fliess@alien-sas.com} \\[.5\baselineskip]
  $^5$ {AL.I.E.N.}, 7 rue Maurice Barr\`es, 54330 V\'ezelise, France \\
  \texttt{} \\[.5\baselineskip]
  $^*$ Corresponding author.
\end{quote}






           



\newpage

\begin{abstract}
  Active Queue Management (AQM) for mitigating Internet congestion has
  been addressed via various feedback control syntheses, especially P,
  PI, and PID regulators, by using a linear approximation where the
  ``round trip time'', i.e., the delay, is assumed to be
  constant. This constraint is lifted here by using a nonlinear
  modeling with a variable delay, introduced more than 20 years
  ago. This delay, intimately linked to the congestion phenomenon, may
  be viewed as a ``flat output.'' All other system variables,
  especially the control variable, i.e., the packet loss ratio, are
  expressed as a function of the delay and its derivatives: they are
  frozen if the delay is kept constant. This flatness-like property,
  which demonstrates the mathematical discrepancy of the linear
  approximation adopted until today, yields also our control strategy
  in two steps: Firstly, designing an open-loop control, thanks to
  straightforward flatness-based control techniques, and secondly,
  closing the loop via Model-Free Control (MFC) in order to take into
  account severe model mismatches, like, here, the number of TCP
  sessions. Several convincing computer simulations, which are easily
  implementable, are presented and discussed.
\end{abstract}


\noindent\textbf{Highlights:}
\begin{itemize}
\item In order to mitigate Internet congestion, this work is among the
  first ones to use a $20$ years old modeling via a nonlinear
  differential equation with a variable delay, where a new
  flatness-like property is encountered: the delay is a flat
  output. Combining flatness-based open-loop control and closed-loop
  control via the intelligent proportional controller deduced from
  model-free control yields easily implementable and convincing
  computer experiments which display a remarkable robustness with
  respect to large uncertainties on  the number of TCP connections.
\item The above nonlinear modeling has mainly been employed until
  today to derive time-invariant linear delay approximate systems,
  which are quite popular, not only for investigating
  control-theoretic questions but also for computer experiments. The
  flatness-like property of the nonlinear model shows that freezing
  the delay implies that all other system variables, including the
  control one, are kept constant. The validity of the linear
  approximations is therefore questioned.
\end{itemize}

\vspace*{\baselineskip}




\section{Introduction}
In order to alleviate Internet congestion, an \emph{active queue
  management} is a dropping packets policy inside a router buffer
yielding a corresponding queue length management (see, e.g.,
\cite{adams}, \cite{varma}, \cite{compar}, \cite{hotchi21} for surveys
and comparisons). It is often related to various control techniques
and should perhaps be viewed, according to \cite{varma}, as ``the
largest human-made feedback-controlled system in the world.'' A
modeling of the most popular \emph{transmission control protocol}
(\emph{TCP}) has been derived more than twenty years ago in
\cite{misra} and \cite{hollot02} via some relationship with fluid
mechanics. It is a nonlinear system of differential equations with a
time-dependent delay, where the control variable is the packet loss
ratio. Although this work is much cited, it seems, to the best of our
knowledge, that almost only linear approximations with constant delays
have been exploited to propose various applicable AQM techniques (see,
however, \cite{barbera,belamfedel,li}). Let us restrict our short
review to a few examples where this approximation has been employed:
\begin{itemize}
\item The familiar \emph{random early detection} (\emph{RED})
  algorithm, which was invented by \cite{red}, has been commented by
  \cite{hollot_red}, \cite{pipd}.
\item The well-known \emph{proportional-integral enhanced} (PIE)
  controller is introduced by \cite{pie}.
\item New algorithms are initiated by \cite{bisoy}, \cite{hotchi1},
  \cite{bottle}, \cite{hotchi21}, \cite{hotchi2}.
\end{itemize}

Our work relies on a remarkable attribute of the above mentioned
nonlinear system: it should be called \emph{flat} in a more or less
analogous sense of \cite{flmr_ijc,flmr_ieee}. The delay is a {\em flat
  output}, the queue length another one. This means that the knowledge
of the delay time variation, or of the queue length, determines all
the other system variables, including the control one.

In particular, freezing the delay implies at once that all the other
system variables are constant. This property thus questions the
frequent use that seemed until today so self-evident, both for
simulation and control purposes, of the linear approximations, where
the delay is assumed to be constant and not the other system variables
(see, e.g., \cite{ALLIOKE}, and references therein).

This paper shows that appropriate control-theoretic tools do exist for
handling the nonlinear modeling:
\begin{itemize}
\item Open-loop control strategies are deduced at once by exploiting
  this flatness-like property of the nonlinear modeling. We here
  choose to regulate the delay:
\begin{itemize}
\item It is obviously related to congestion.
\item \emph{Controlling queue Delay} (\emph{CoDel}) (\cite{jacobson}),
  which has become a popular setting, also puts delay control on the
  forefront but via a completely different viewpoint.
\end{itemize}
\item In order to counteract the unavoidable model mismatches (see,
  e.g., \cite{xu} for a summary of the shortcomings of the above
  nonlinear modeling) and disturbances, we follow \cite{villagra},
  \cite{cancer}, \cite{covid} by closing the loop via
  \emph{model-free} control in the sense of \cite{csm1,csm2}. The
  presence of a delay requires a predictor which cannot, here, be the
  celebrated Smith's predictor (\cite{smith}), because the latter is
  model-based (see, e.g., \cite{plestan} for a recent survey). We thus
  adapt here the viewpoint of \cite{supply} for studying \emph{supply
    chain management} (see also \cite{marseille}). This is achieved by
  removing the unpredictable \emph{quick fluctuations} via a theorem
  due to \cite{cartier} which is expressed in the language of
  \emph{nonstandard analysis}. Note that this result has led to a new
  understanding of time series (see, e.g, \cite{solar}, and references
  therein).
\end{itemize}

Our paper is organized as follows. After showing that the delay or the
queue length can be viewed as a flat output, Section \ref{nlmodel}
explains the inherent weakness of linear approximations with a
constant delay. Section \ref{nodelay} recalls the basic facts of
model-free control, which has already been used successfully many
times. In order to take the delay into account in the model-free
approach, Section~\ref{delay} exploits techniques stemming from
nonstandard analysis. Numerical simulations are presented in Section
\ref{simu}: they show that the model mismatch on the number of TCP
sessions is well compensated by the closed-loop control without any
clear-cut superiority of the techniques developed in Section
\ref{delay}. Various questions are raised in Section \ref{conclusion}.

\section{Some consequences of the nonlinear modeling}\label{nlmodel}
\subsection{Flatness}\label{flat}

The nonlinear TCP/AQM network model (\cite{misra,hollot02}) reads
\begin{subequations}\label{nl}
\begin{align}
    \label{nl1}
 \dot{W}(t) &= \frac{1}{R(t)} -  \frac{W(t)W(t - R(t))}{2[R(t - R(t))]}u(t - R(t)) \\
\label{nl2}
\dot{Q}(t) &= \frac{W(t)}{R(t)} N(t) - C(t)  
\end{align}
\end{subequations}
where
\begin{itemize}
\item $W(t) > 0$ is the length of the TCP window;
\item $R(t)> 0$ is the \emph{round trip time} (\emph{RTT}) which
  appears as a time-dependent delay in Equation \eqref{nl};
\item $Q(t) > 0$ is the queue length;
\item the control variable is the dropping packet policy $u(t)$,
  $0 \leqslant u(t) \leqslant 1$: it is called the \emph{packet loss
    ratio} (\cite{ALLIOKE}), or, as often in the literature, the
  \emph{packet drop probability};
\item $C(t) > 0$ is the bottleneck link capacity;
\item $N(t) > 0$ is the number of TCP sessions. It plays the role of
  external disturbance.
\end{itemize}
In practice $C(t)$ and $N(t)$ are piecewise constant. Thus
$\dot{C}(t) = \dot{N}(t) = 0$, with the exception of a finite number
of points on any finite-time interval. The RTT $R(t)$ and the queue
length $Q(t)$ are related by a simple affine relation
\begin{equation}\label{affine}
 R(t) = T + \frac{Q(t)}{C(t)}  
\end{equation}
where $T$ is the round trip propagation time. Assume, in the following
computations, $\dot{C}(t) = \dot{N}(t) = 0$ on some time interval, Set
$C(t) = C$, $N(t) = N$, where $C$ and $N$ are constant.  Then Equation
\eqref{affine} yields $\dot{Q}(t) = C \dot{R}(t)$. Equation
\eqref{nl2} becomes
\begin{equation}\label{nl3}
\dot{R}(t) = \frac{N W(t)}{C R(t)} - 1  
\end{equation}
It yields
\begin{equation}\label{flat1}
    W(t) = \frac{C R(t) (\dot{R}(t) + 1)}{N}
\end{equation}
Thus Equations \eqref{flat1} and \eqref{nl1} show that $W(t)$ and
$u(t - R(t))$ depend on $R(t)$ and its first and second order
derivatives. In other words, we may call System \eqref{nl} \emph{flat}
and the RTT $R(t)$ a \emph{flat output} (compare with
\cite{mounier}). Equation \eqref{affine} shows that the queue length
$Q(t)$ is another flat output.
\begin{rmk}
  Classic flatness has been formally defined in \cite{flmr_ijc} via
  differential algebra and in \cite{flmr_ieee} via differential
  geometry of infinite jets and prolongations. Combining differential
  and difference algebras (see, e.g., \cite{cohn}) permits a precise
  definition of flatness for nonlinear systems with constant delays
  (\cite{mounier_rudolph1,mounier_rudolph2}).  Such a setting does not
  however work with a variable delay such as $R(t)$, since the time
  derivation $\frac{d}{dt}$ and the time shift $t \mapsto t - R(t)$
  with a time-varying quantity do not commute. Before developing an
  adequate general mathematical formalism for our example, it may be
  wise to wait for other concrete case-studies. It would open a path
  to a new understanding of nonlinear systems with variable delays.
\end{rmk}

\subsection{Critical appraisal of the linear approximation}\label{linear}
See \cite{ALLIOKE} for a nice survey on linear approximations. Let
$u_0$, $Q_0$, $W_0$, $R_0$ be the numerical values of $u(t)$, $Q(t)$,
$W(t)$, $R(t)$ at an operating, or equilibrium, point. Contrarily to
the variables $u$, $Q$, $W$, the delay $R$ is kept frozen at the value
$R_0$: the delay in the linear approximation is constant. It is
obvious that such an assumption contradicts Section \ref{flat}, where
Equations \eqref{affine}, \eqref{flat1} and \eqref{nl1} show that
$Q(t)$, $W(t)$ and $u(t)$ become constant when $R(t)$ is
constant. This fact is casting some doubt not only about AQM via such
approximations, but also on the computer simulations, which rely on it
(see \cite{ali,ALLIOKE}, and the references therein).

\begin{rmk}
Define the control variable $\delta u (t) = u(t) - u_0$ and the output variable $\delta Q (t) = Q (t) - Q_0$. They are often related in the literature (see, e.g., \cite{ALLIOKE}) by the time-invariant linear delay system defined by the transfer function
\begin{equation*}\label{transf}
  - \frac{ (2N\frac{W_0}{2})^3 e^{- R_0 s}}{(R_0 s + 1)(\frac{W_0 R_0}{2} s + 1)}
  \end{equation*}
where the number $N$ of sessions is assumed to be constant. A system defined by such a transfer function is sometimes called  
\emph{quasi-finite} (\cite{marquez}). The output $\delta Q$ is said  to be {\em flat}, or \emph{basic} (\cite{marquez}). 
\end{rmk}

\section{Closed-loop control via model-free control without delay: a short review}\label{nodelay}
\subsection{Ultra-local model}
Consider a single-input single-output (SISO) nonlinear system ($\Sigma$). Denote by $\mathfrak{u} (t)$ (resp. $\mathfrak{y} (t)$) the control (resp. output) variable.
It has been demonstrated (\cite{csm1}) via elementary techniques from functional analysis and differential algebra that the often poorly known modeling of ($\Sigma$) may be replaced, if some quite weak assumptions are satisfied, by an \emph{ultra-local model}:
\begin{equation}
\mathfrak{y}^{(\nu)} = F + \alpha \mathfrak{u} \label{ul}
\end{equation}
where $\alpha \in \mathbb{R}$ is chosen by the practitioner such that $\alpha\mathfrak{u}$ and $\mathfrak{y}^{(\nu)}$  are of the same order of magnitude: it does not need to be precisely known.
Numerous successful applications (see, e.g., references in \cite{csm1,csm2}) have shown that $\nu = 1$ in Equation \eqref{ul} yields most often a convenient ultra-local model:
\begin{equation}\label{1}
\dot{\mathfrak{y}} = F + \alpha \mathfrak{u}
\end{equation}
The following comments are useful:
\begin{itemize}
\item Equation \eqref{1} is only valid during a short time lapse: it must be continuously updated.
\item $F$ is \emph{data-driven}, i.e., it is estimated via the knowledge $\mathfrak{u}$ and $\mathfrak{y}$ (\cite{csm1}):
\begin{equation}\label{integral}
F_{\text{est}}(t)  =-\frac{6}{\tau^3}\int_{t-\tau}^t \left\lbrack (\tau -2\sigma){\mathfrak{y}}(\sigma)+\alpha\sigma(\tau -\sigma){\mathfrak{u}}(\sigma)\right \rbrack d\sigma 
\end{equation}
The quantity $\tau > 0$ may be chosen to be quite ``small.'' The above integral, which is a low pass filter, may, 
in practice, be  replaced by a classic digital filter.
\item $F$ subsumes not only the unknown structure of the system, {which most of the time is nonlinear}, but also 
any external disturbance.
\end{itemize}

\subsection{Intelligent controllers and local stability}
The loop is closed with the following \emph{intelligent proportional controller} (\cite{csm1}), or \emph{iP},
\begin{equation}\label{ip}
\mathfrak{u} = - \frac{F_{\text{est}} - \dot{\mathfrak{y}}^\star + K_P e }{\alpha}
\end{equation}
where:
\begin{itemize}
\item $\mathfrak{y}^\star$ is the reference trajectory of the output,
\item $e = \mathfrak{y} - \mathfrak{y}^\star$ is the tracking error,
\item $K_P \in \mathbb{R}$ is a tuning gain.
\end{itemize}
Combining Equations \eqref{1} and \eqref{ip} yields
$\dot{e} + K_P e = F - F_{\text{est}}$. If the estimate $F_{\text{est}}$ is ``good,'' i.e., if $F - F_{\text{est}} \approx 0$, then $\lim_{t \to + \infty} e(t) \approx 0$, if, and only if, $K_P > 0$.

\section{Closed-loop control via model-free control with delay}
\label{delay}
\subsection{Ultra-local model with delay}
Set $v(t) = u(t - R(t))$ in Equation \eqref{nl}:
\begin{subequations}
\begin{align*}
 \dot{W}(t) &= \frac{1}{R(t)} -  \frac{W(t)W(t - R(t))}{2[R(t - R(t))]}v(t) \\
\dot{Q}(t) &= \frac{W(t)}{R(t)} N(t) - C(t)  
\end{align*}
\end{subequations}
This trivial change of variable shows that the techniques from \cite{csm1} remain valid for introducing the ultra-local model with a time-varying delay
\begin{equation}\label{1R}
\dot{\mathfrak{y}}(t) = \mathfrak{F} + \alpha \mathfrak{u}(t - R(t))
\end{equation}
where $\mathfrak{F}$ plays the same r\^{o}le as $F$ in Equation \eqref{1}. Equation \eqref{integral} becomes
\begin{equation}\label{integralR}
\mathfrak{F}_{\text{est}}(t)  =-\frac{6}{\tau^3}\int_{t-\tau}^t \left\lbrack (\tau -2\sigma)\mathfrak{y}(\sigma)+\alpha\sigma(\tau -\sigma)\mathfrak{u}(\sigma - R(\sigma) \right\rbrack d\sigma 
\end{equation}

\subsection{Prediction via time 
series}\label{tool}
\subsubsection{Time series and the Cartier-Perrin theorem}
Consider the time
interval $[0, 1] \subset \mathbb{R}$. Introduce as often in
\emph{nonstandard analysis} (see \cite{robinson}, \cite{diener}, \cite{lobry}) the infinitesimal sampling of $[0,1]$: ${\mathfrak{T}} = \{ 0 = t_0 < t_1 < \dots < t_\nu = 1 \}$
where $t_{i+1} - t_{i}$, $0 \leqslant i < \nu$, is {\em infinitesimal},
{\it i.e.}, ``very small''. A
\emph{time series} $X(t)$ is a function $X: {\mathfrak{T}}
\rightarrow \mathbb{R}$.

A time series ${\mathcal{X}}: {\mathfrak{T}} \rightarrow \mathbb{R}$
is said to be {\em quickly fluctuating}, or {\em oscillating}, if 
the integral $\int_A {\mathcal{X}} dm$ is
infinitesimal, \textit{i.e.}, very small, for any \emph{appreciable} interval, \textit{i.e.}, an interval which is neither ``very small'' nor ``very large''.

According to a theorem due to
\cite{cartier}, the following additive decomposition holds for any time series~$X$, which satisfies a weak integrability condition,
\begin{equation}\label{decomposition}
X(t) = E(X)(t) + X_{\tiny{\rm fluctuation}}(t)
\end{equation}
where
\begin{itemize}
\item the \emph{mean}, or \emph{trend}, $E(X)$ is ``quite smooth'';
\item $X_{\tiny{\rm fluctuation}}$ is quickly fluctuating.
\end{itemize}
The decomposition \eqref{decomposition} is unique up to an additive
infinitesimal: It means that the two terms on the right handside of Equation \eqref{decomposition} are unique up to a ``very small'' additive quantitity.

\subsubsection{Derivative estimate}

Let us start with a polynomial time function of degree $1$
$$
p_{1} (\tau) = a_0 + a_1 \tau  
$$
where $\tau \geqslant 0$, $a_0, a_1 \in \mathbb{R}$. Operational calculus (see, \textit{e.g.}, \cite{yosida}) with respect to the variable~$\tau$, permits to express  $p_1$ as 
$$P_1 = {a_0}/{s} + {a_1}/{s^2}$$ 
Multiply both sides by $s^2$:
\begin{equation}\label{5}
s^2 P_1 = a_0 s + a_1
\end{equation}
Take the derivative of both sides with respect to $s$, which
corresponds in the time domain to the multiplication by $- \tau$:
\begin{equation}\label{6}
s^2 \frac{d P_1}{ds} + 2s P_1 = a_0
\end{equation}
The coefficients $a_0, a_1$ are obtained via the triangular system
of linear equations (\ref{5})-(\ref{6}). We get rid of the time
derivatives, \textit{i.e.}, of $s P_1$, $s^2 P_1$, and $s^2 \frac{d
P_1}{ds}$, by multiplying both sides of Equations
(\ref{5})-(\ref{6}) by $s^{ - n}$, $n \geq 2$. The corresponding
iterated time integrals are low pass filters which attenuate the
corrupting noises. A quite short time window is sufficient for
obtaining accurate values of $a_0$, $a_1$.
\begin{rmk}
See \cite{mboup} and \cite{othmane1,othmane2} for more details. Note also that estimating derivatives via integrals seems to have been first introduced by \cite{lanczos}.
\end{rmk}

\subsubsection{Prediction}\label{predic}
Set the following forecast $X_{\text{forecast}}(t + \Delta T)$, where $\Delta T > 0$ is not too ``large'',
\begin{equation}\label{forecast}
X_{\text{forecast}}(t + \Delta T) = E(X)(t) + \left[\frac{d E(X)(t)}{dt}\right]_e \Delta T
\end{equation}
where $E(X)(t)$ and $\left[\frac{d E(X)(t)}{dt}\right]_e$ are estimated like $a_0$ and $a_1$ above. Let us stress that what we predict is the mean and not the quick fluctuations.
\begin{rmk}
The above construction is obviously reminiscent of the \emph{sliding window} techniques in the applied literature on time series (see, e.g., \cite{melard}). 
\end{rmk}
 Note that estimating $a_0$ and $a_1$ yields respectively the mean and the derivative.
 
\subsubsection{Local closed-loop stability}
Equation \eqref{1R} may be rewritten as
\begin{equation}\label{1S}
\dot{\mathfrak{y}}(t + S(t)) = \mathfrak{F}_{\text{forecast}}(t + S(t)) + \alpha \mathfrak{u}(t)
\end{equation}
where the advance $S(t) > 0$ is defined by
\begin{equation}\label{advance}
    S(t) = \min \left\{ \,\tau\ |\ \tau - R(\tau) = t\, \right\} - t
\end{equation}
If $R(t)$ is ``slowly'' varying, it is clear that $R(t)$ and $S(t)$ remain close. Evaluating $S(t)$ however requires a prediction of $R(t)$. Replace therefore in Equation \eqref{advance} $R(t)$ by the reference trajectory $R^\star (t)$. It yields 
\begin{equation}\label{adv}
   S^\star(t) = \min \{ \tau | \tau - R^\star(\tau) = t \} - t 
\end{equation}
Equation \eqref{ip} then becomes
\begin{equation}\label{ipR}
\mathfrak{u}(t) = - \frac{\mathfrak{F}_{\text{forecast}}(t + S^\star(t)) - \dot{\mathfrak{y}}^\star (t + S^\star(t)) + K_P e{(t + S^\star(t))} }{\alpha}
\end{equation}

where:
\begin{itemize}
\item the forecast of $\mathfrak{F}$ is obtained via Formulae \eqref{integralR}  and \eqref{forecast};
\item $e (t + S^\star(t)) = \mathfrak{y}_{\text{forecast}}(t + S^\star(t)) - \mathfrak{y}^\star (t + S^\star(t))$, where $\mathfrak{y}^\star$ is the reference trajectory and $e$ is the tracking error;
\item $\mathfrak{y}_{\text{forecast}}(t + S^\star(t)) = \mathfrak{z}(t + S^\star(t))$ is obtained via the linear differential equation $$\dot{\mathfrak{z}}(\tau) = \mathfrak{F}_{\text{forecast}}(\tau) + \alpha \mathfrak{u}(\tau -R^\star(\tau)) \quad t \leqslant \tau \leqslant t + S(t)$$

\item $K_P \in \mathbb{R}$ is the tuning gain.
\end{itemize}
It yields
\begin{align}
    \label{errBF}
\dot{e}(t + S^\star(t)) + K_P e(t + S^\star(t)) = \mathfrak{F} (t + S^\star(t)) - \mathfrak{F}_{\text{forecast}}(t + S^\star(t))
\end{align}
Local stability is ensured, i.e., $\lim_{t \to +\infty} \mathfrak{y} (t + S^\star(t)) \approx \mathfrak{y}_{\text{forecast}}(t + S^\star(t))$ , if 
\begin{itemize}
\item $K_P > 0$,
\item the forecast is ``good,'' i.e., ${\mathfrak{F} (t + S^\star(t)) - \mathfrak{F}_{\text{forecast}}(t + S^\star(t))} \approx 0$.
\end{itemize}
%

\section{Computer simulations\protect\footnote{Contact C. Join ({\tt cedric.join@univ-lorraine.fr}) for the simulation codes.}}\label{simu}
\subsection{Various situations}
Introduce the following control settings:

\begin{enumerate}
    \item {\bf Reference trajectory and nominal control}: The choice of a reference trajectory $R^\star(t)$ for the delay $R(t)$ yields at once via Section \ref{flat} an open-loop nominal control $u^\star(t)$ for $u(t)$, i.e.,
    $$
    u^\star(t-R^\star(t))=2\left(\frac{1}{R^\star(t)}-\dot W^\star(t)\right)\left(\frac{R^\star(t-R^\star(t))}{W^\star(t) W^\star(t-R^\star(t))}\right)
    $$
or
$$
    u^\star(t)=2\left(\frac{1}{R^\star(t+S^\star(t))}-\dot W^\star(t+S^\star(t))\right)\left(\frac{R^\star(t)}{W^\star(t+S^\star(t)) W^\star(t)}\right)
    $$
and
$$
    W^\star (t) = \frac{C(t) R^\star(t) (\dot{R^\star}(t) + 1)}{N_0}
$$
where $N_0 = 60$ and $C=3000$. 

\item {\bf Open-loop control (OL)}: Inject $u^\star$ in Equation \eqref{nl} with $N(t) = N_0 $.

\item {\bf Closing the loop via an iP without delay (iP)}: In Equations \eqref{1}, \eqref{integral}, \eqref{ip}, set $\mathfrak{y}(t) = e(t) = R(t) - R^\star (t)$, $\mathfrak{y}^\star (t) = 0$, $\mathfrak{u}(t) = \Delta u (t) = u(t) - u^\star (t)$. The iP \eqref{ip} becomes
\begin{equation*}\label{ipsr}
\Delta u = - \frac{F_{\text{est}} + K_P e }{\alpha}
\end{equation*}
where $\alpha = -1000$, \textcolor{black}{$K_P = 1$}. Consider the estimation of $F_{\text{est}}$ via Formula \eqref{integral} as a classic \emph{finite impulse response} (\emph{FIR}) (see, e.g., \cite{fir}). We thus apply a control $u(t)$ of the form
$$ 
  u(t) = u^\star (t) - \frac{F_{\text{est}} + K_P e }{\alpha}
$$
where the first part incorporates our knowledge of the system, and
the second one deals with perturbations, model imperfections and
unknown dynamics. It is clear that the practical implementation, which has already been achieved successfully a number of times, is  straightforward.

\item {\bf Closing the loop via an iP with delay (iPWD)}: In Equations \eqref{1R}, \eqref{integralR}, \eqref{ipR}, set as above $\mathfrak{y}(t) = e(t) = R(t)- R^\star (t)$, $\mathfrak{y}^\star (t) = 0$, $\mathfrak{u}(t) = \Delta u (t) = u(t) - u^\star (t)$.
The iP \eqref{ipR} becomes
$$
\Delta u (t) = - \frac{\mathfrak{F}_{\text{forecast}}(t + S^\star(t)) + K_P e(t+S^\star(t)) }{\alpha}
$$
where, as above, $\alpha = -10$, \textcolor{black}{$K_P = 1$}. The calculations related to predictions are detailed in \cite{solar}. The sequel is similar to the case without delay.
\end{enumerate}


\subsection{Scenarios}\label{sce}
\begin{figure*}[!ht]
\centering%
\subfigure[\footnotesize Control (--) and nominal control (- -) ]
{\epsfig{figure=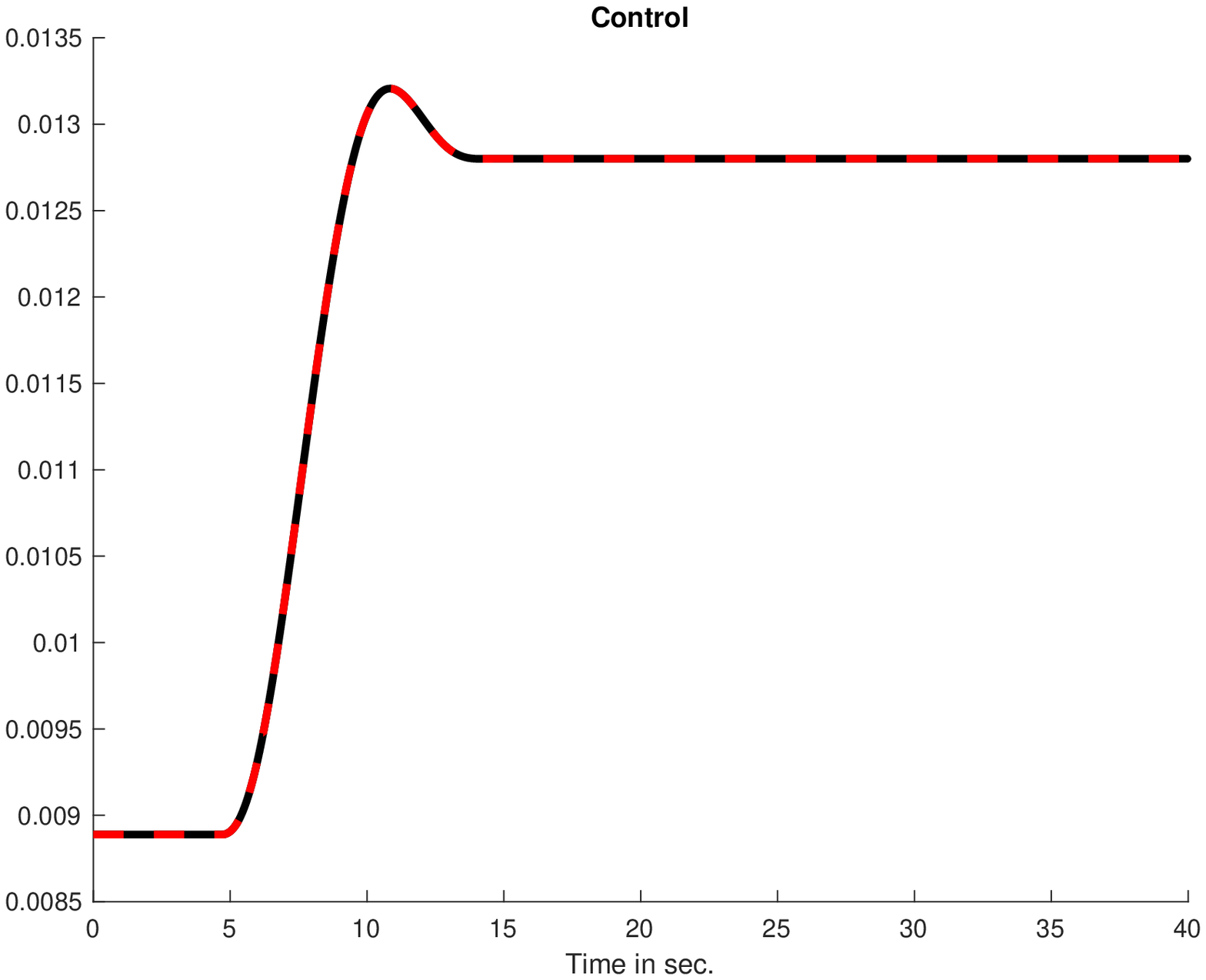,width=0.48\textwidth}}
\subfigure[\footnotesize $R$ (--) and reference trajectory (- -)]
{\epsfig{figure=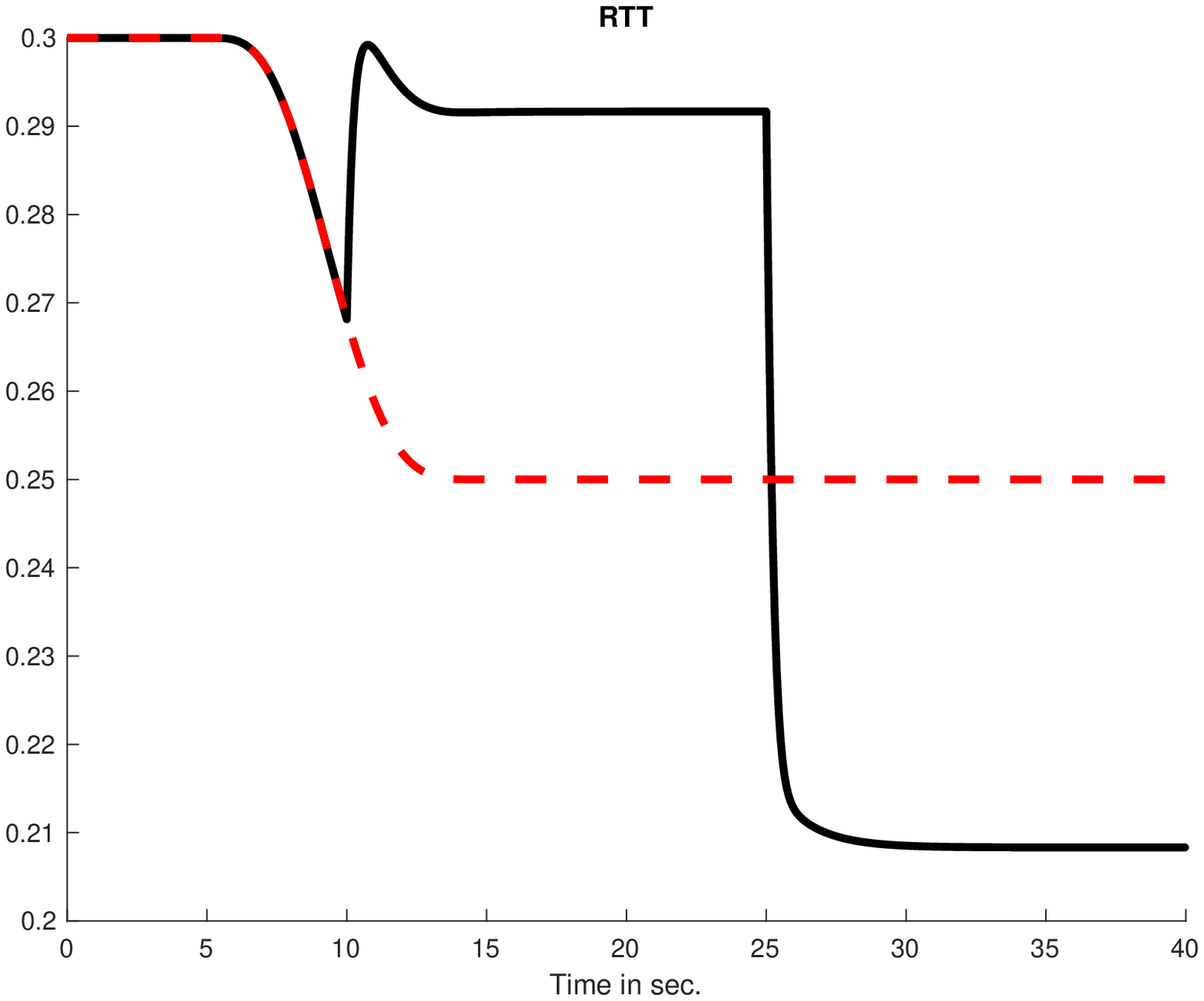,width=0.48\textwidth}}
\\
\subfigure[\footnotesize TCP Window]
{\epsfig{figure=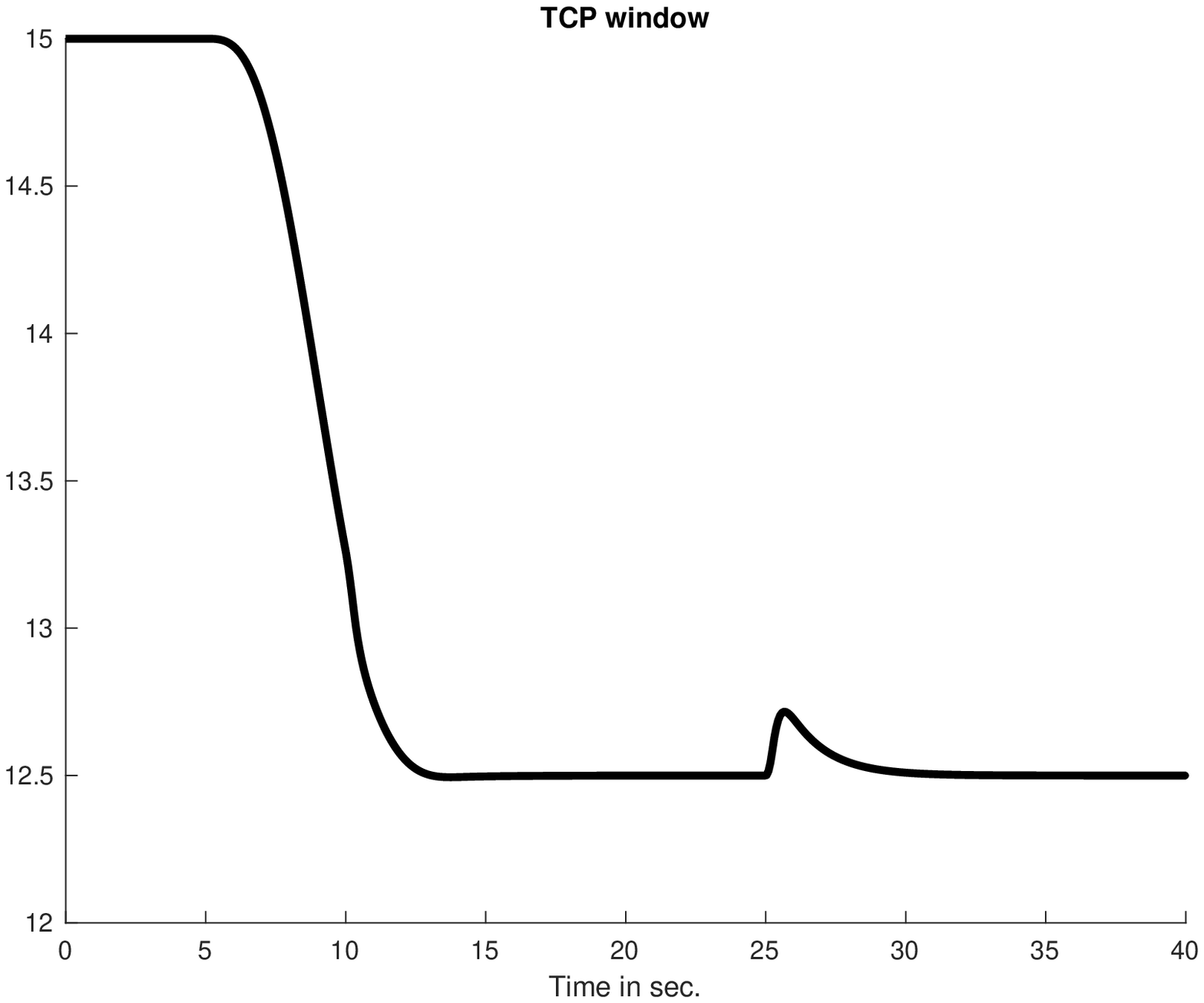,width=0.33\textwidth}}
\subfigure[\footnotesize Queue length]
{\epsfig{figure=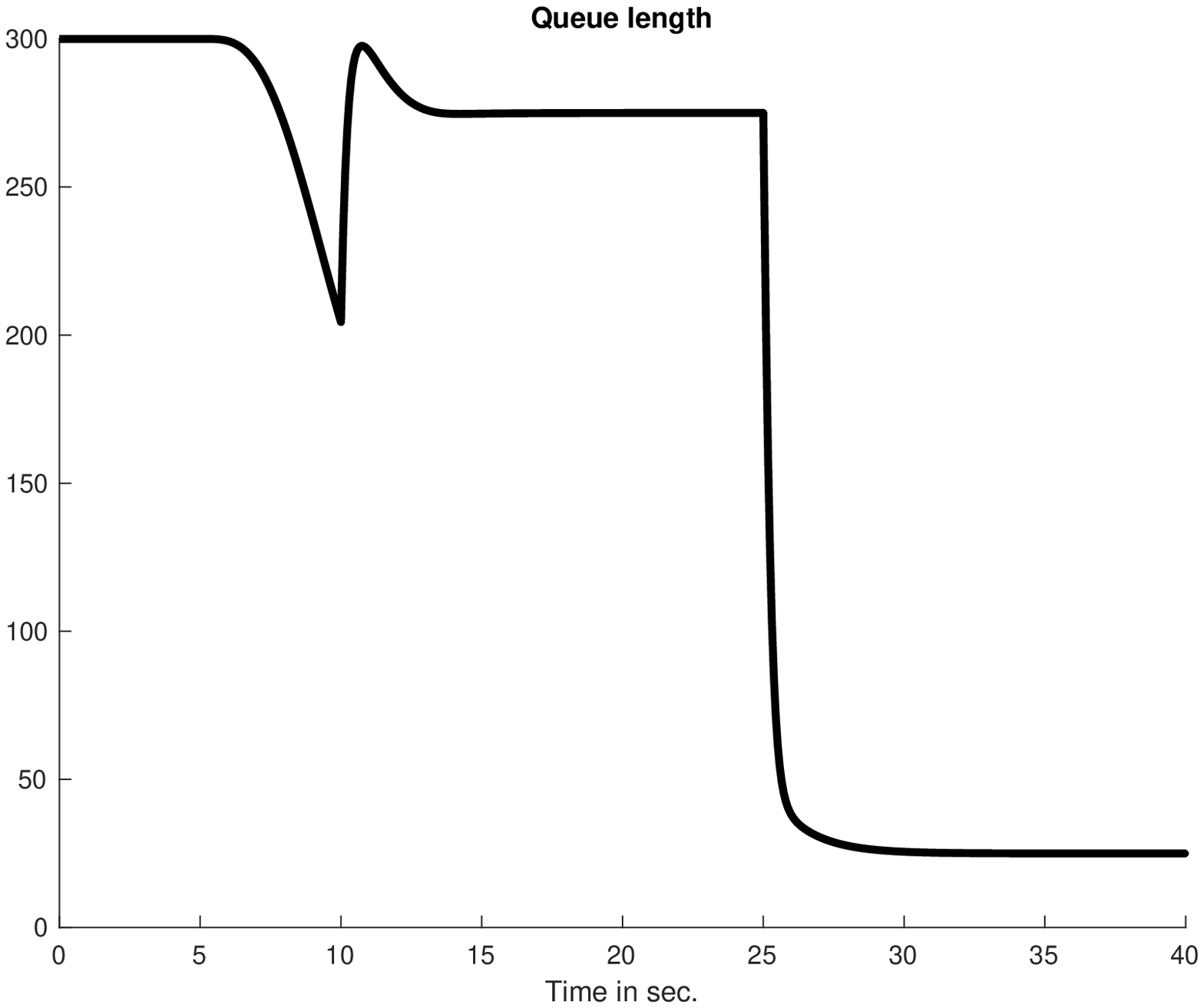,width=0.33\textwidth}}
\subfigure[\footnotesize Number of connections]
{\epsfig{figure=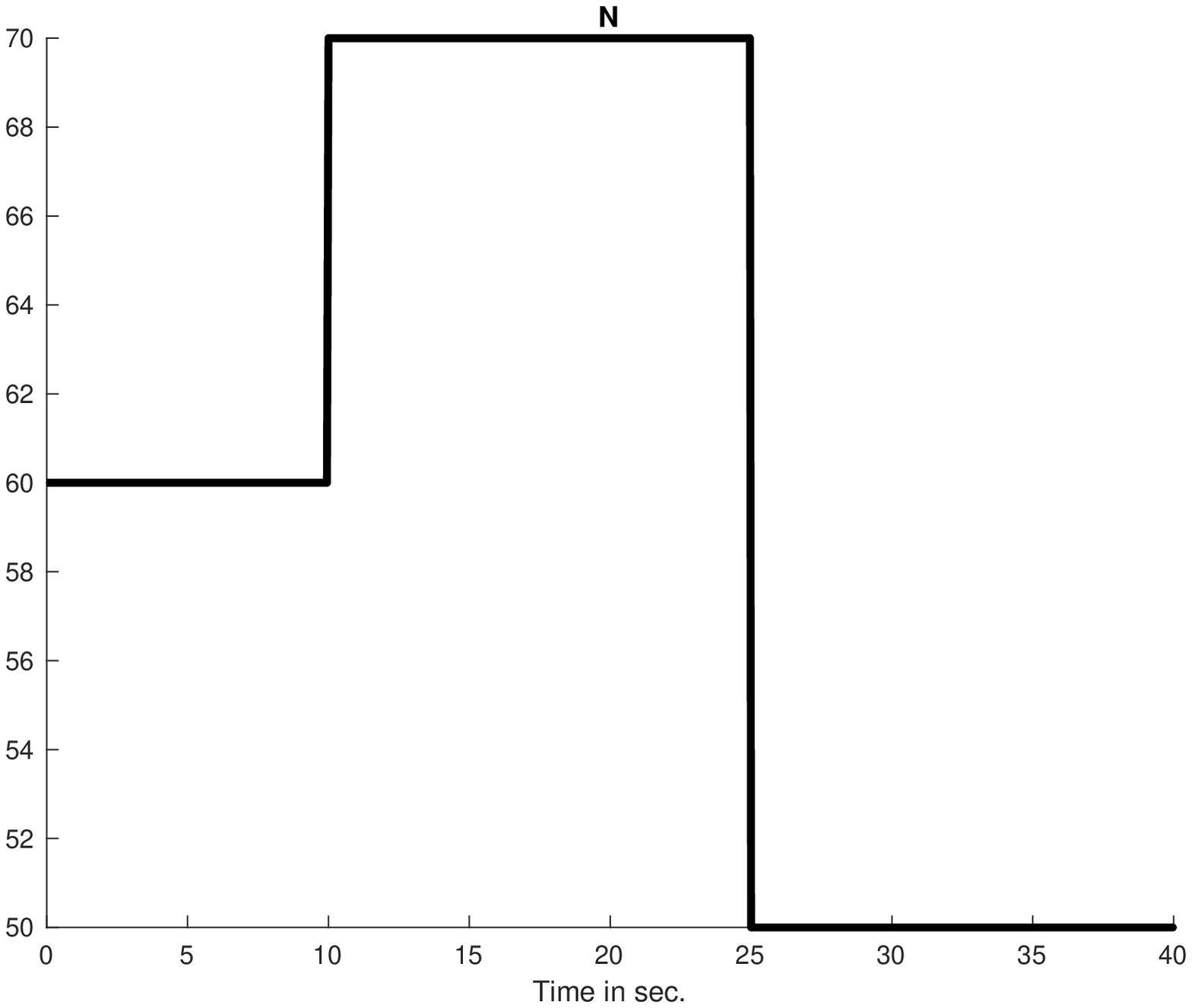,width=0.33\textwidth}}
\caption{Scenario 1 -- OL}\label{S20}
\end{figure*}

\begin{figure*}[!ht]
\centering%
\subfigure[\footnotesize Control (--) and nominal control (- -) ]
{\epsfig{figure=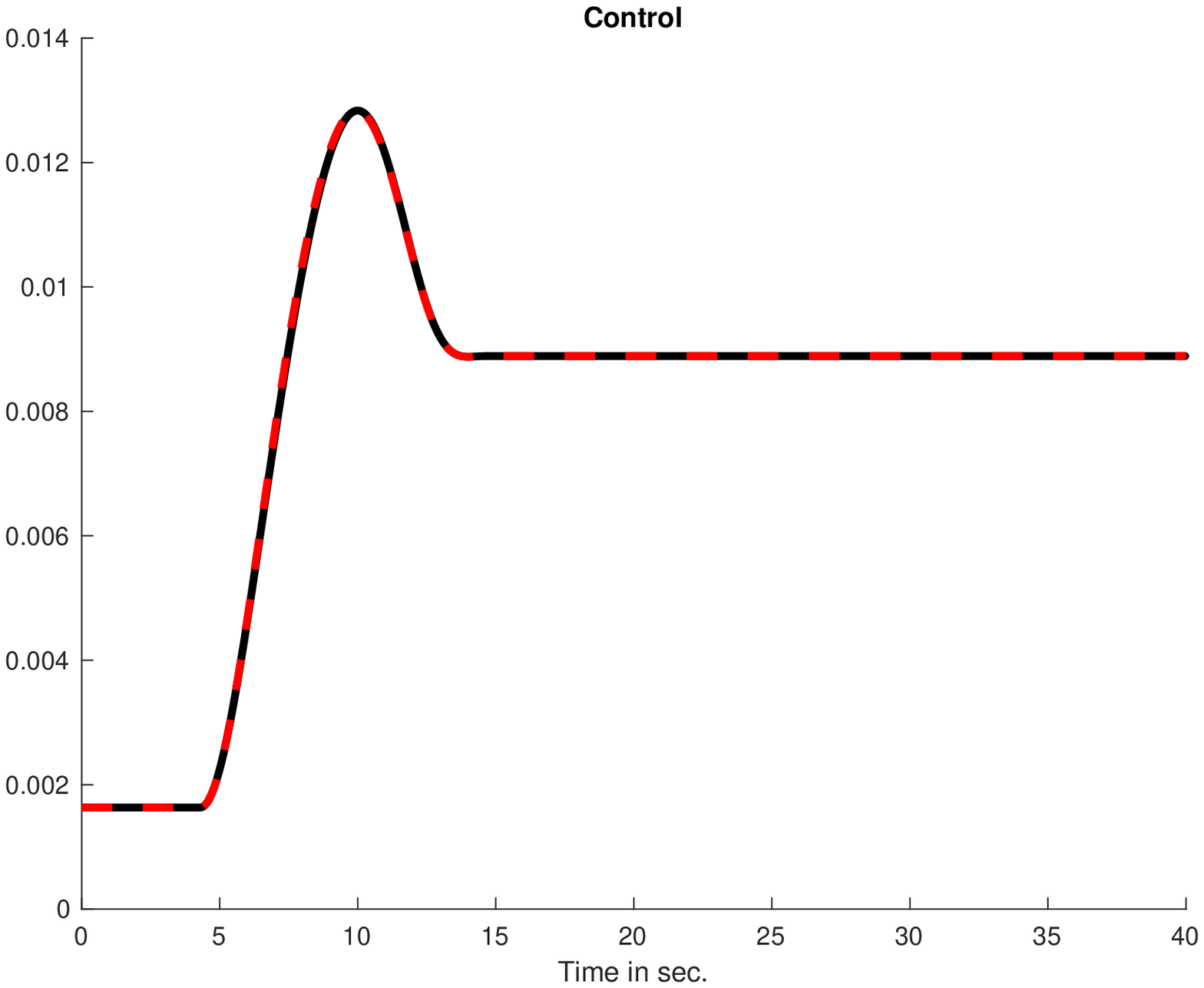,width=0.48\textwidth}}
\subfigure[\footnotesize $R$ (--) and reference trajectory (- -)]
{\epsfig{figure=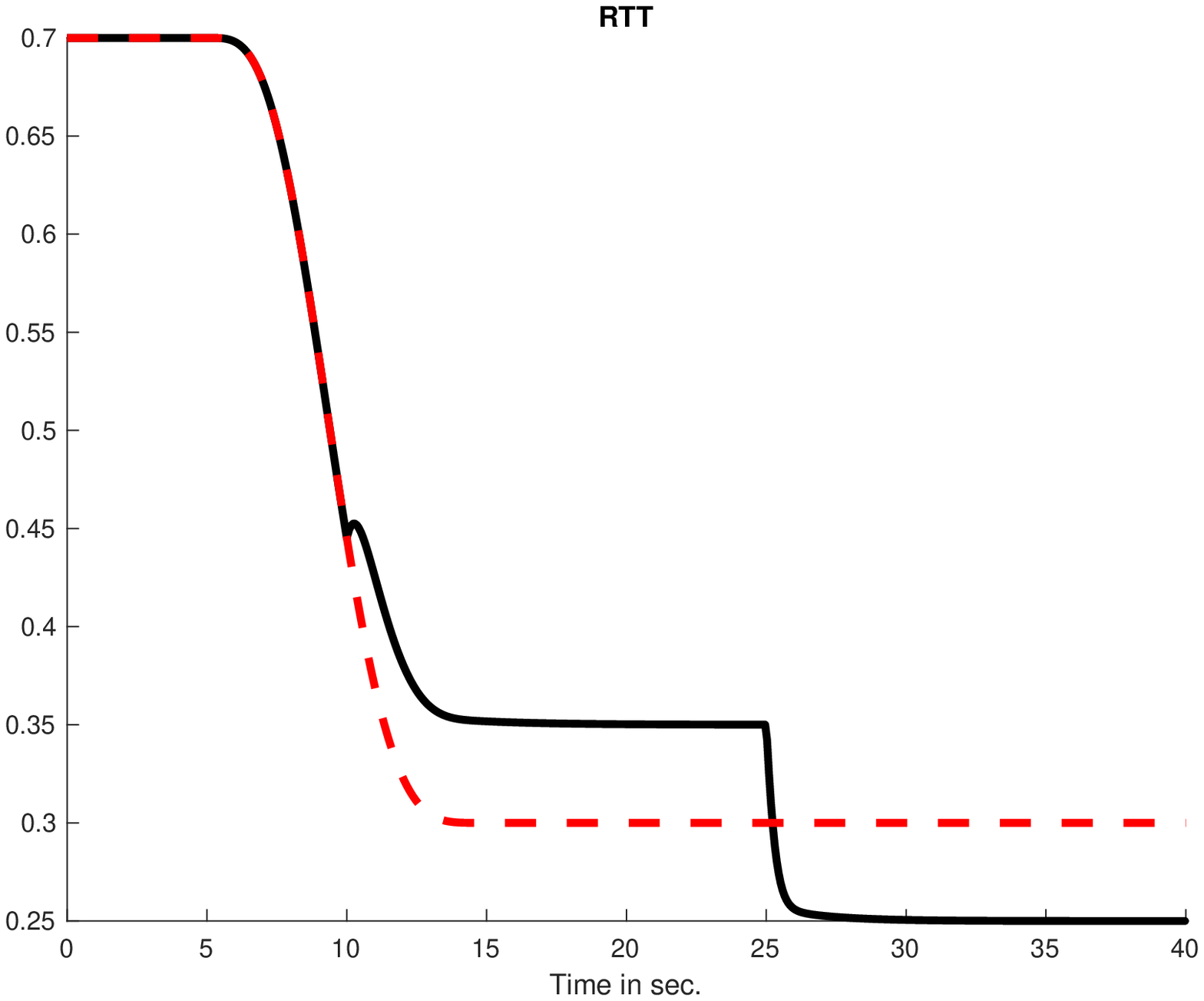,width=0.48\textwidth}}
\\
\subfigure[\footnotesize TCP Window]
{\epsfig{figure=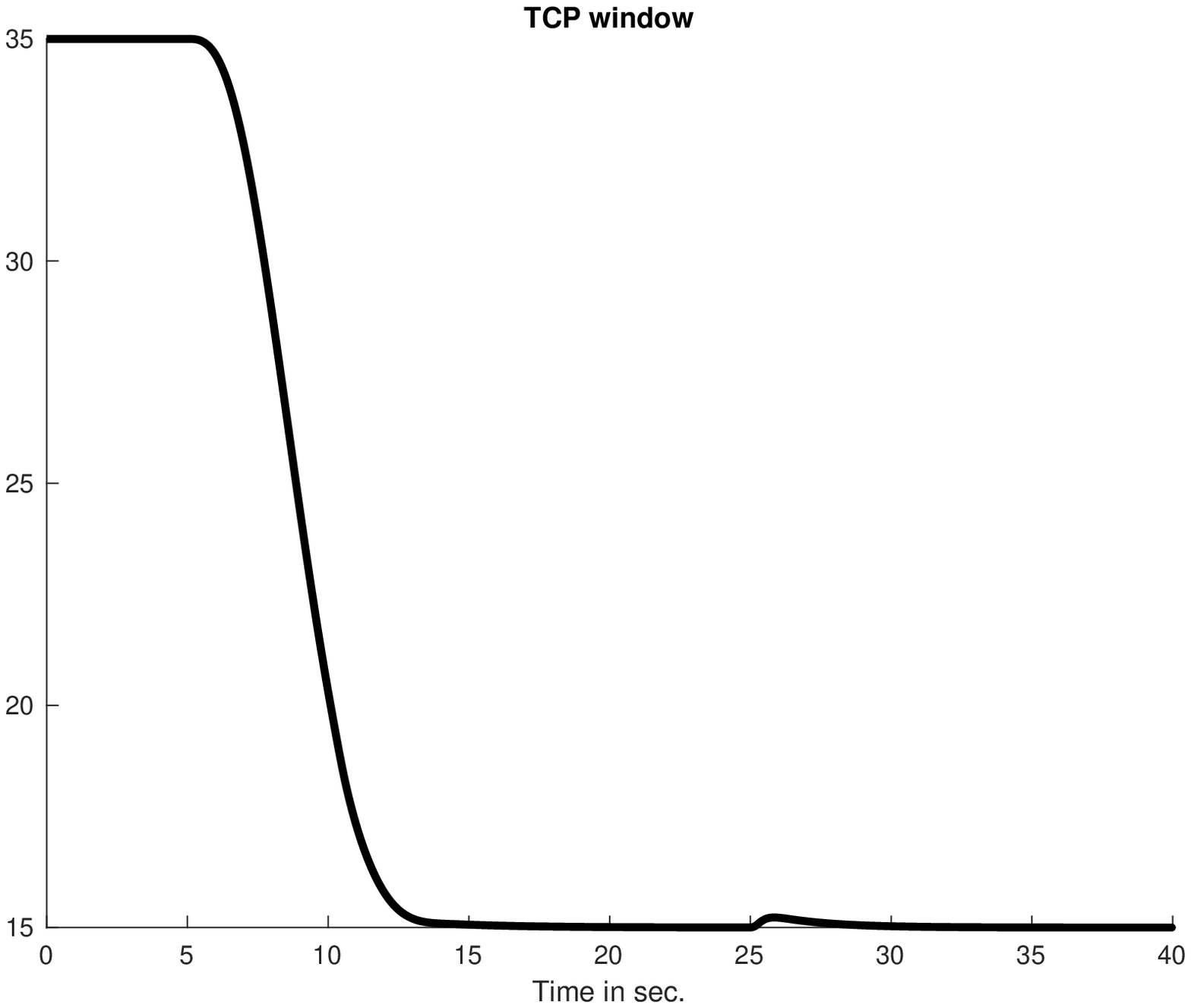,width=0.33\textwidth}}
\subfigure[\footnotesize Queue length]
{\epsfig{figure=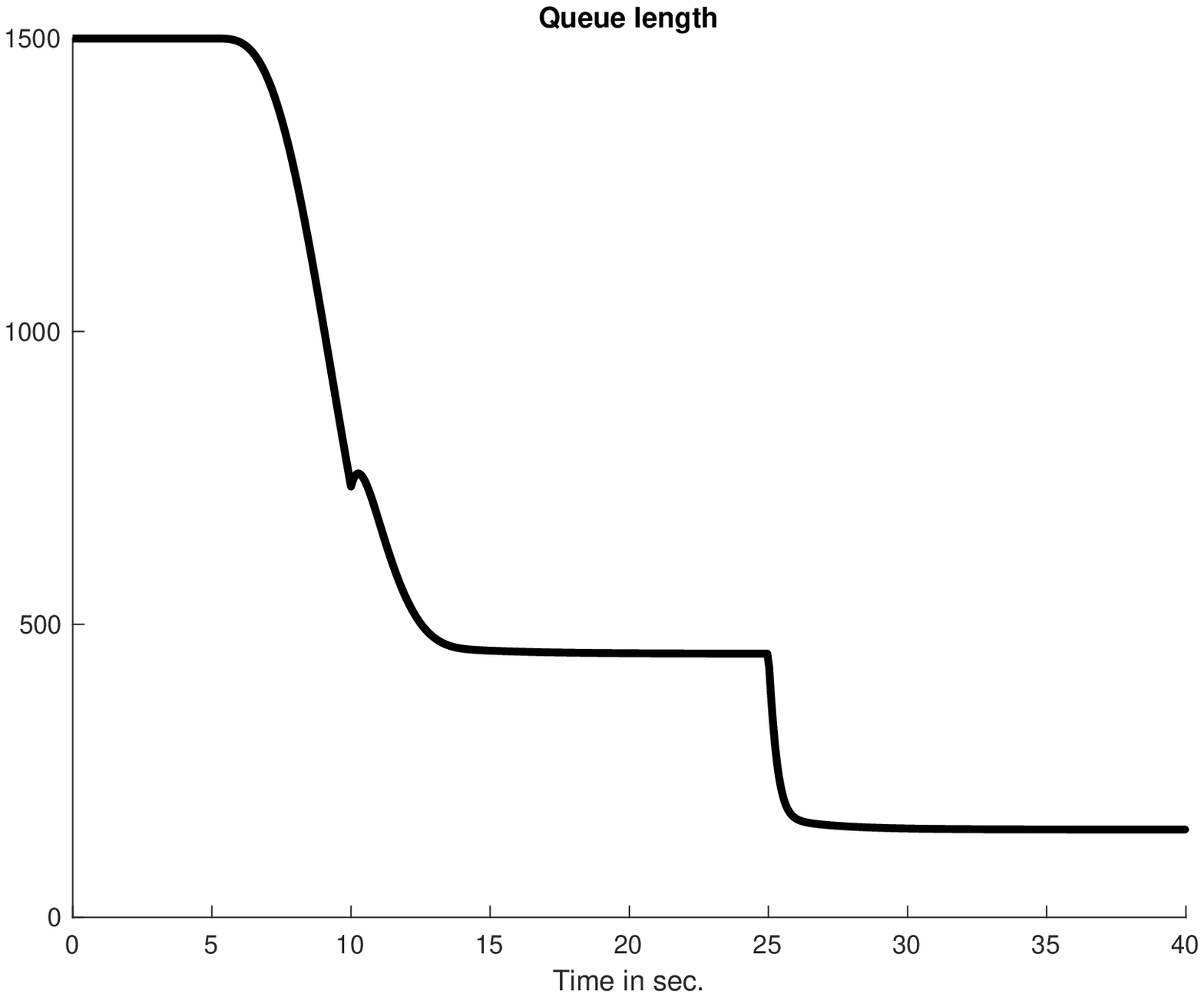,width=0.33\textwidth}}
\subfigure[\footnotesize Number of connections]
{\epsfig{figure=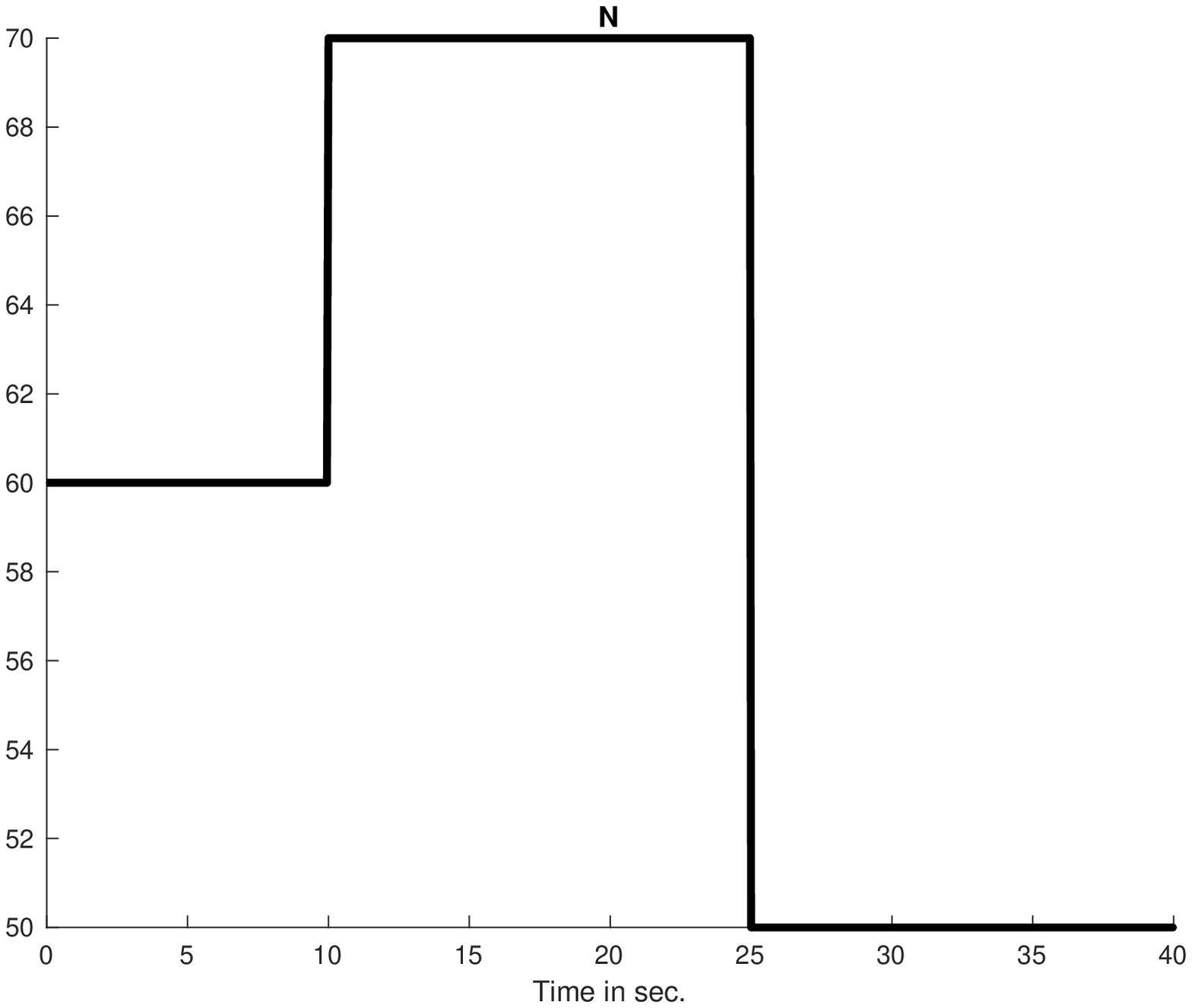,width=0.33\textwidth}}
\caption{Scenario 2 -- OL}\label{S40}
\end{figure*}

We illustrate our control laws through two different scenarios, where the mismatch is the number of TCP sessions. The first scenario corresponds to normal operation: small variation of R. The second scenario represents an exit from a congestion situation and corresponds to a large variation of R. The command is designed with a known value of the number of connections N but operates with a N constant piecewise: in our simulation N first goes from 60 to 70 then goes down to 50. The number of connection N plays the role of an external disturbance. It is moreover this effect that the open loop curves Figs.\ref{S20} and~\ref{S40} show where the trajectory deviates from the reference one when the number of connections changes.
\begin{enumerate}
    \item Scenario $1$: $0.25$s $\leqslant  R(t) \leqslant 0.3$s, $50 \leqslant N(t) \leqslant 70$.
     \item Scenario $2$: $0.3$s $\leqslant  R(t) \leqslant 0.7$s, $50 \leqslant N(t) \leqslant 70$.
\end{enumerate}

\begin{table}[h!]
\begin{center}
\caption{Simulations and Figures}\label{tb}
\begin{tabular}{|c|ccc|}
\hline
Scenarios & OL & iP & iPWD \\\hline\hline
1 & Figure \ref{S20} &  Figure \ref{S22}  &Figure \ref{S24}\\
2&Figure \ref{S40} & Figure \ref{S42}  &Figure \ref{S44} \\\hline
\end{tabular}
\end{center}
\end{table}

\begin{figure*}[!ht]
\centering%
\subfigure[\footnotesize Scenario 1 iP (--), iPWD (- .) and zero line (- -) ]
{\epsfig{figure=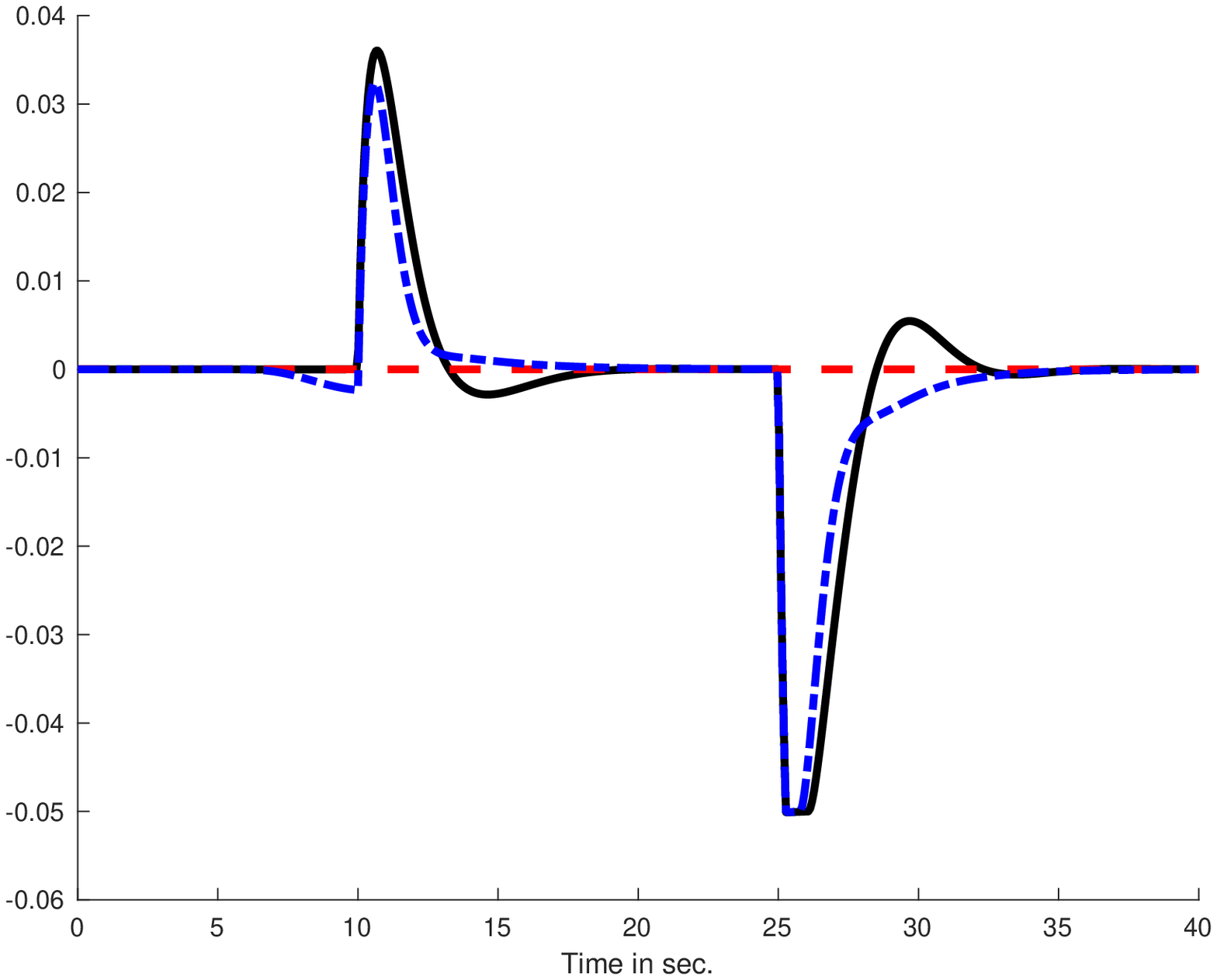,width=0.48\textwidth}}
\subfigure[\footnotesize Scenario 2 iP (--), iPWD (- .) and zero line (- -) ]
{\epsfig{figure=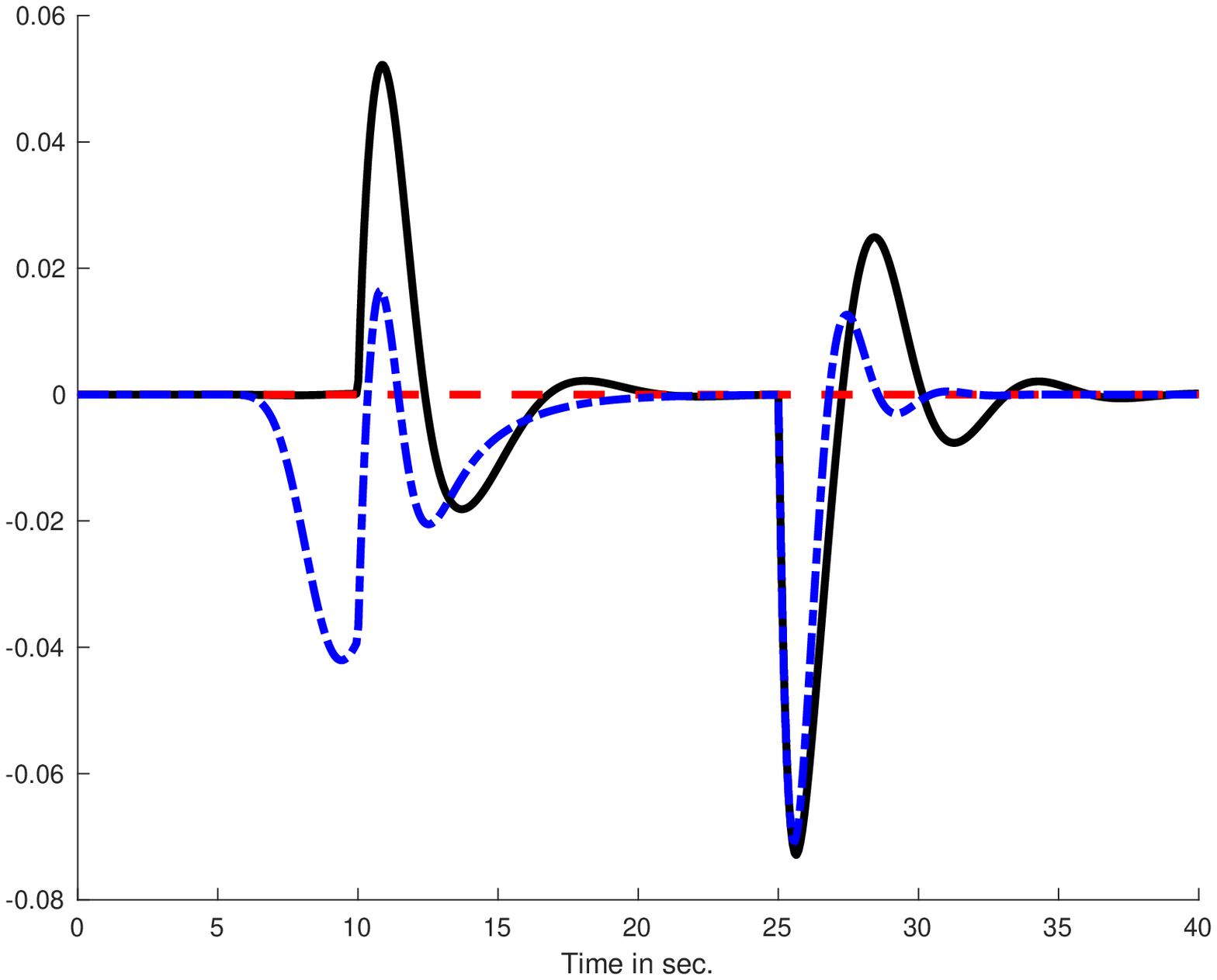,width=0.48\textwidth}}
\caption{Tracking errors}\label{SR}
\end{figure*}

Those Figures tell us that:
\begin{itemize}
\item the mismatch is well compensated by iPs with or without delay,
\item iPs with or without delay exhibit very similar behaviors.
\end{itemize} 
Our results are especially well displayed in Figure \ref{SR}. Indeed, we can see that the behavior of the tracking errors are totally comparable. In other words the iP with delay seems to be useless!

\section{Conclusion}\label{conclusion}
It has been shown that
\begin{itemize}
    \item the linear constant delay approximations, which also play a key r\^{o}le in computer simulations, contradict the more complete nonlinear modeling;

\item control-theoretic tools are available for an active queue management via this nonlinear modeling.
\end{itemize}
Many points remain of course to be addressed:
\begin{itemize}
 \item Other mismatches and external disturbances ought to be examined: noisy measurements, abrupt changes of the round trip time, ... Would the \emph{intelligent proportional-derivative controller} (\emph{iPD}) advocated by \cite{csm2} be helpful? See, e.g., \cite{pd} and \cite{pipd} for results with classic PD controllers.
    \item The simulations in Section \ref{sce} indicate the futility of an iP with delay in order to compensate a model mismatch. Without a precise mathematical analysis, it is not clear whether this  property is always valid. Let us suggest nevertheless that the open-loop nonlinear control, where the delay is taken into account, is doing the job!
    \item The coefficient $\alpha$ in Equations \eqref{1} and \eqref{1R},  which does not need to be determined precisely, is obtained via trials and errors. A more subtle estimation would be welcome. Let us also add that important variations of some quantities like the number $N(t)$ of TCP connections might necessitate the introduction of a time-varying $\alpha$ (see \cite{brest,madrid} for first results in other engineering domains).
    \item Many network simulation for investigating Internet congestion (see, e.g., \cite{riley}, \cite{ali,ALLIOKE}, and references therein) seem to have employed time-invariant linear delay systems (see Section \ref{linear}). It should therefore be most rewarding to develop and integrate the tools of this paper. 
\end{itemize}

\begin{figure*}[!ht]
\centering%
\subfigure[\footnotesize Control (--) and nominal control (- -) ]
{\epsfig{figure=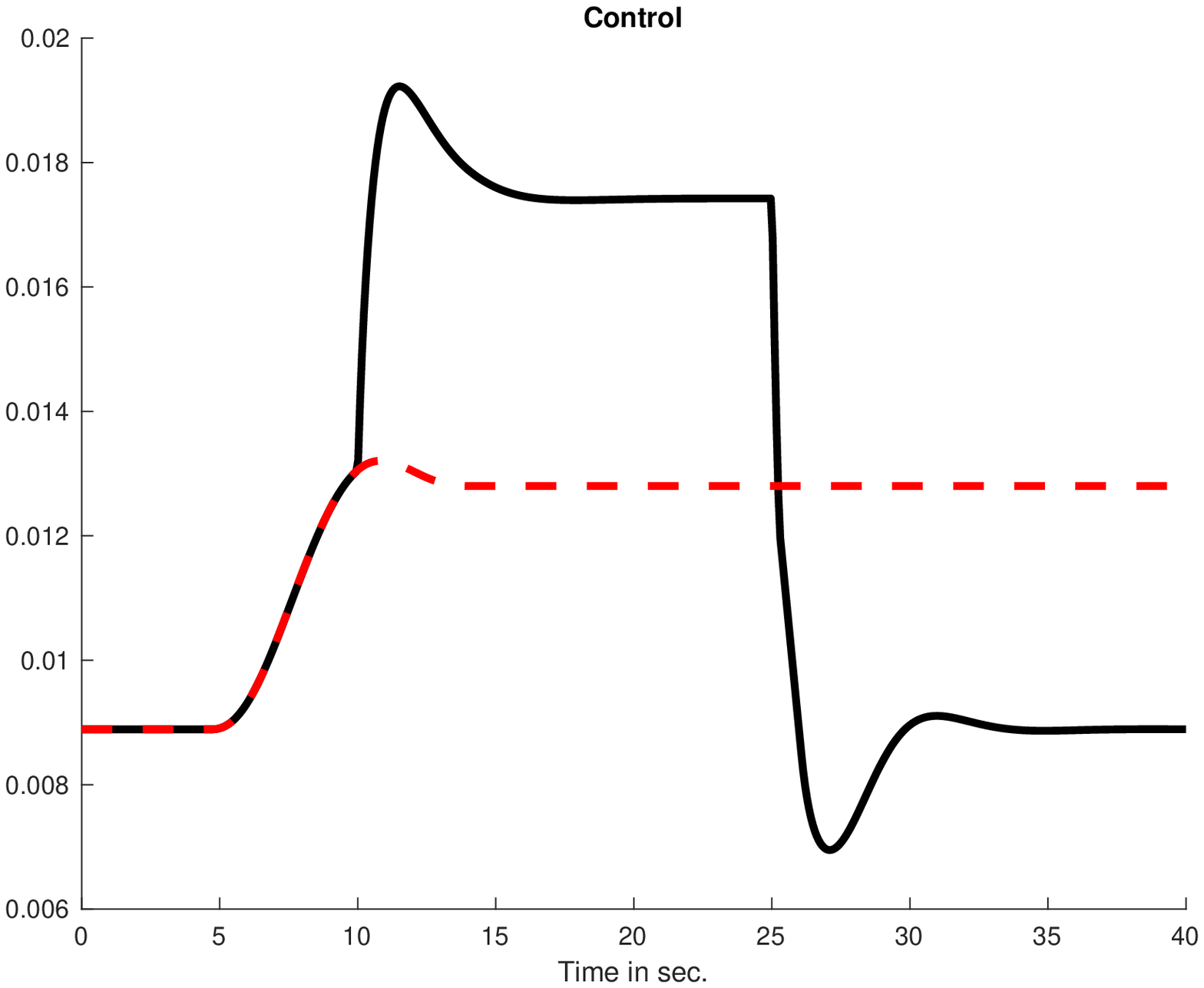,width=0.48\textwidth}}
\subfigure[\footnotesize $R$ (--) and reference trajectory (- -)]
{\epsfig{figure=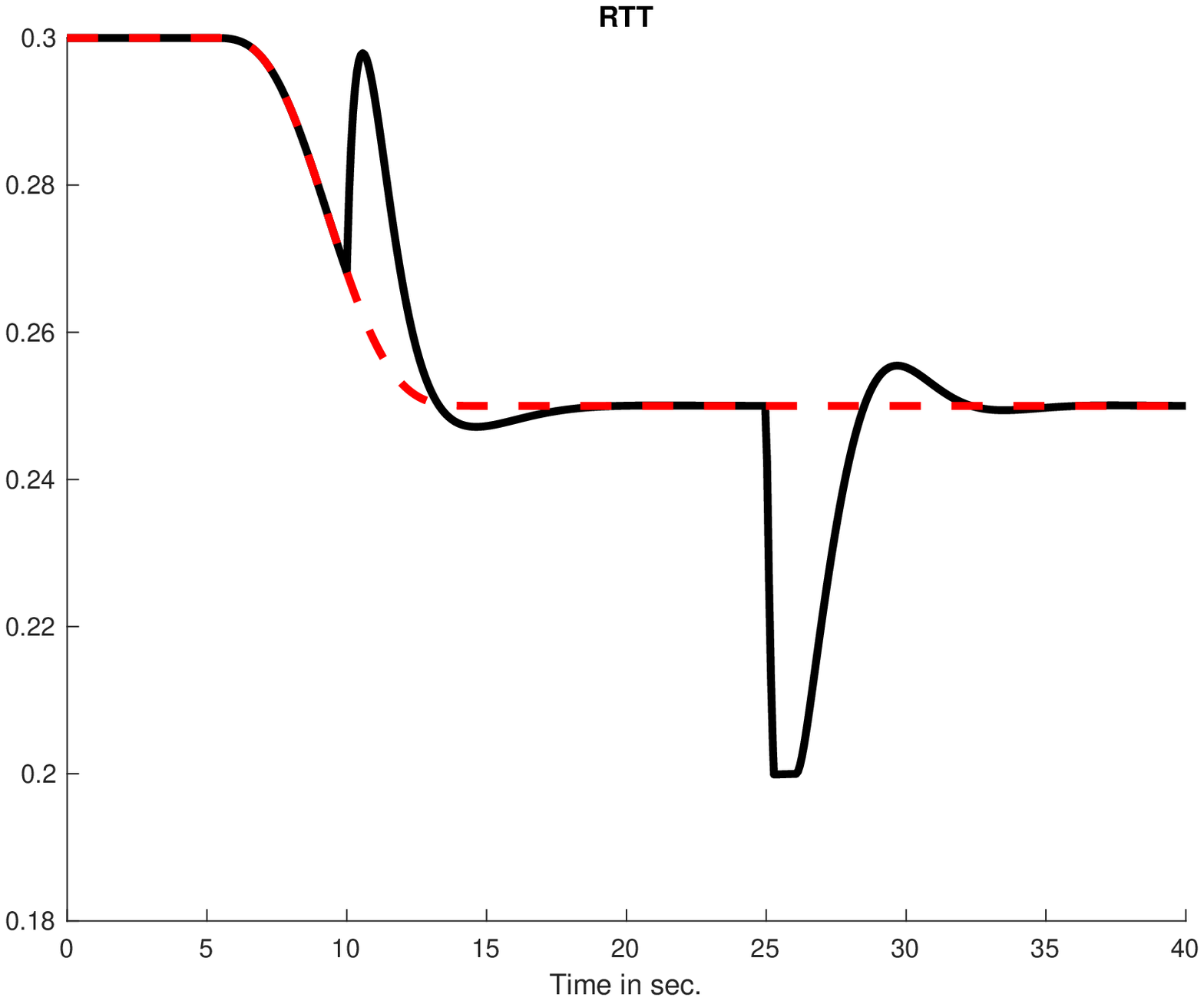,width=0.48\textwidth}}
\\
\subfigure[\footnotesize TCP Window]
{\epsfig{figure=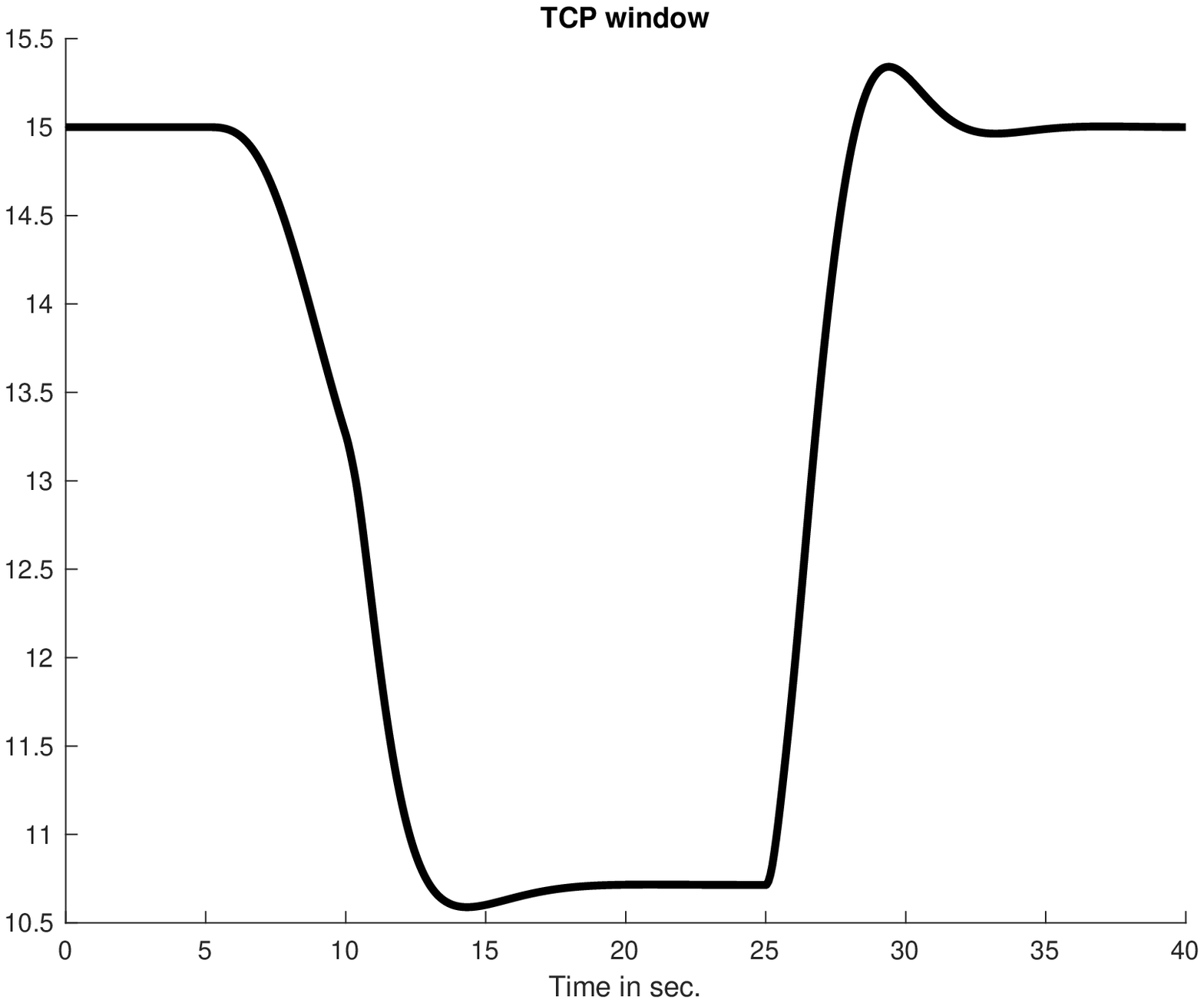,width=0.33\textwidth}}
\subfigure[\footnotesize Queue length]
{\epsfig{figure=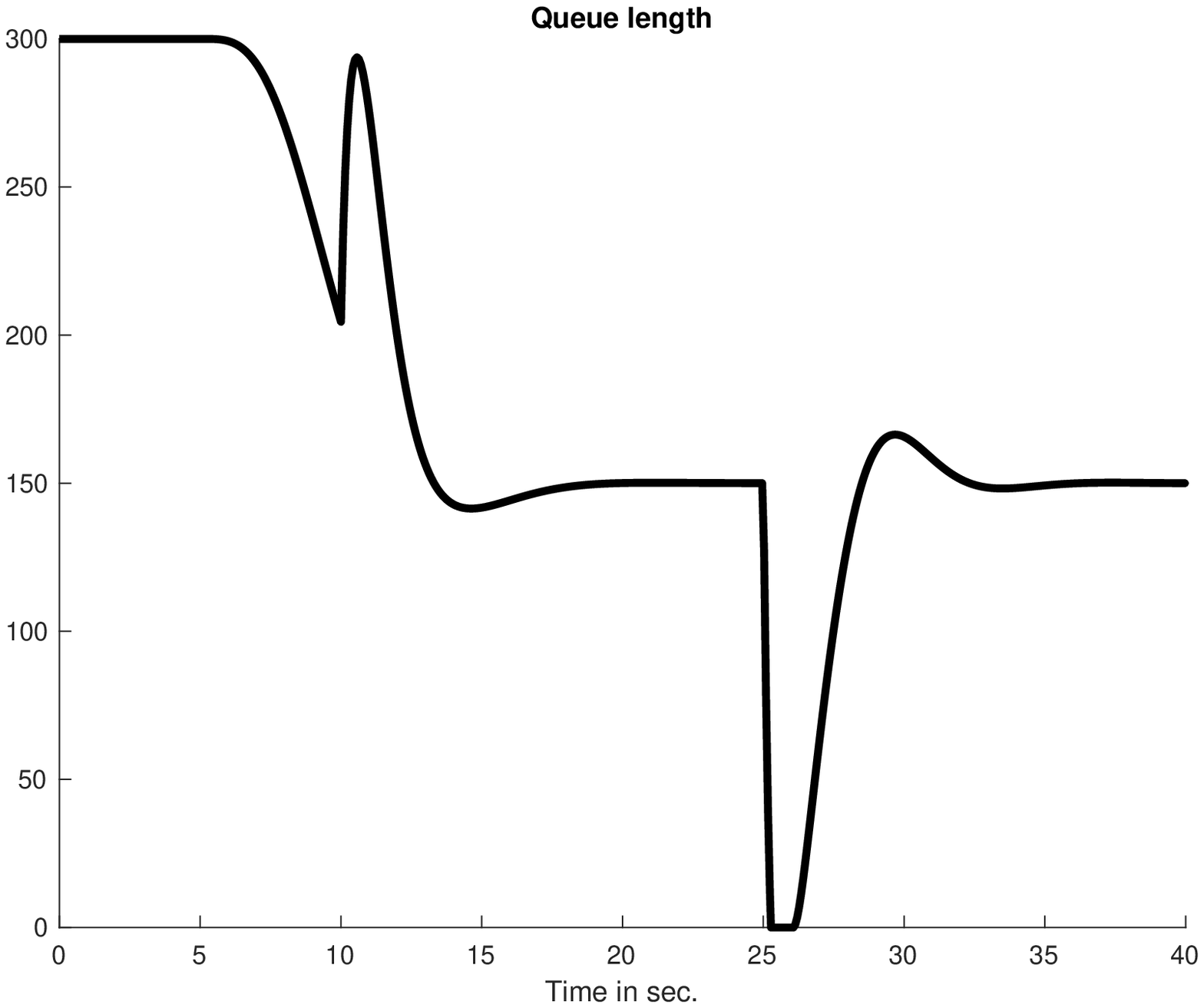,width=0.33\textwidth}}
\subfigure[\footnotesize Number of connections]
{\epsfig{figure=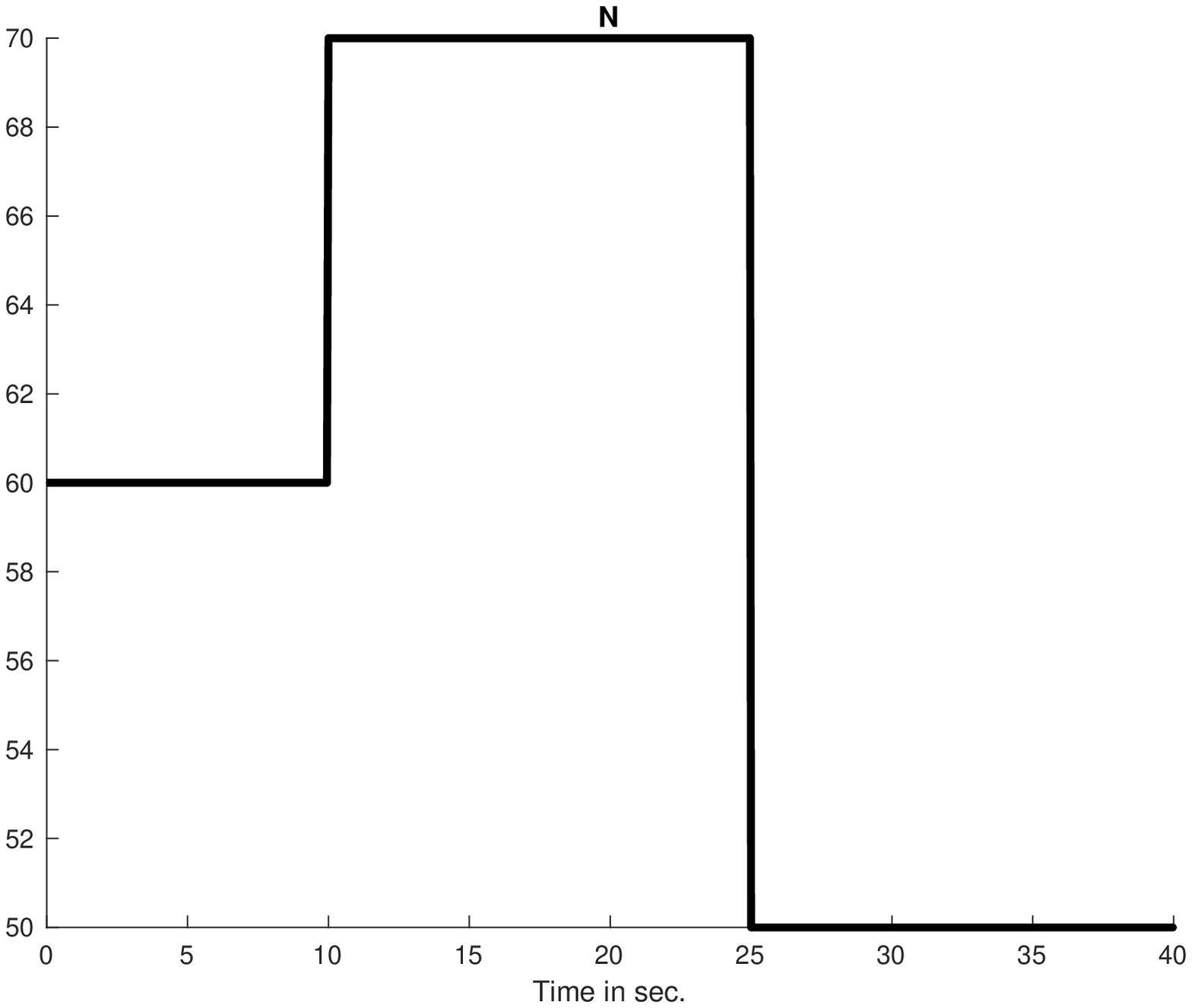,width=0.33\textwidth}}
\caption{Scenario 1 -- iP}\label{S22}
\end{figure*}


%
\begin{figure*}[!ht]
\centering%
\subfigure[\footnotesize Control (--) and nominal control (- -) ]
{\epsfig{figure=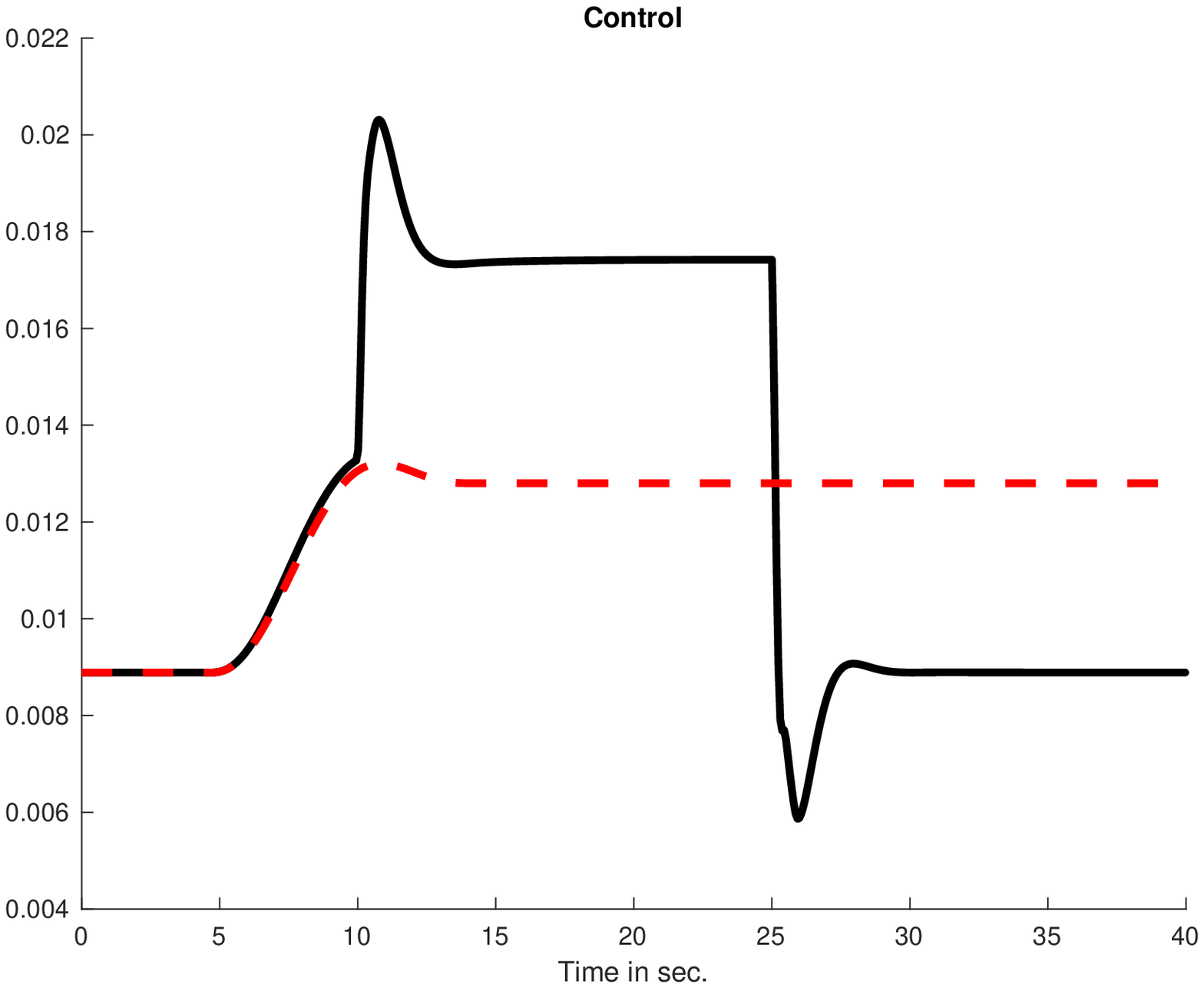,width=0.48\textwidth}}
\subfigure[\footnotesize $R$ (--) and reference trajectory (- -)]
{\epsfig{figure=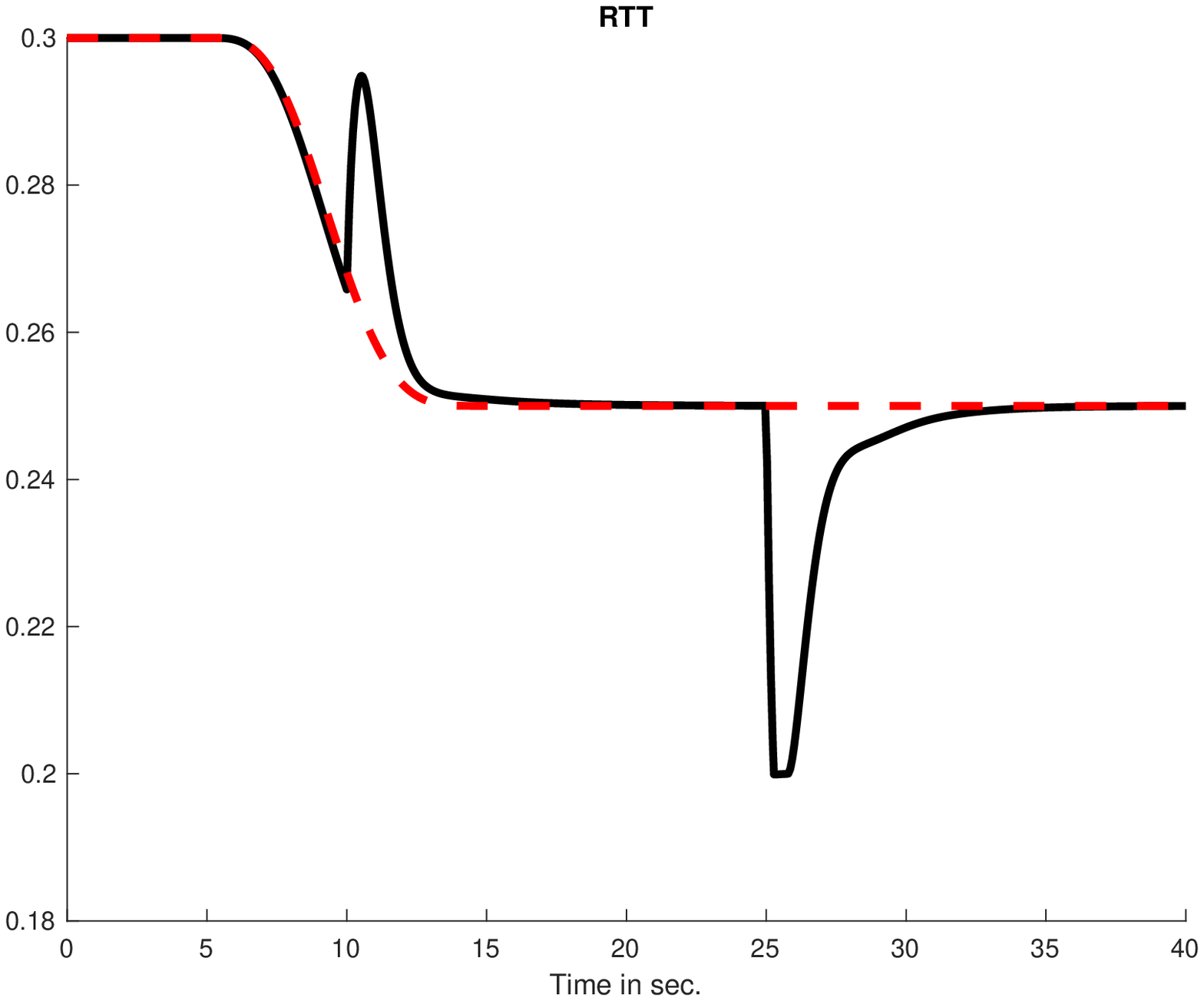,width=0.48\textwidth}}
\\
\subfigure[\footnotesize TCP Window]
{\epsfig{figure=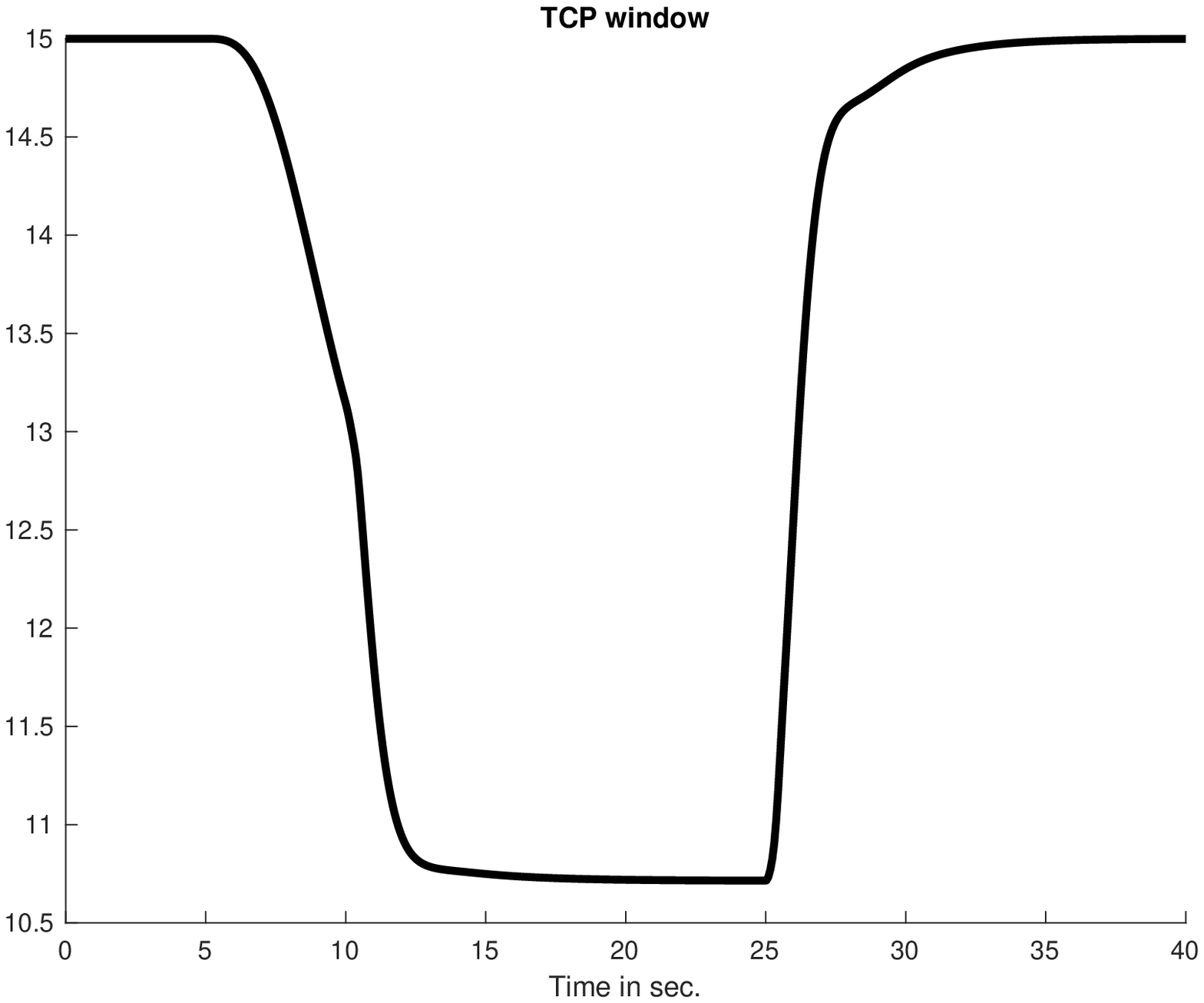,width=0.33\textwidth}}
\subfigure[\footnotesize Queue length]
{\epsfig{figure=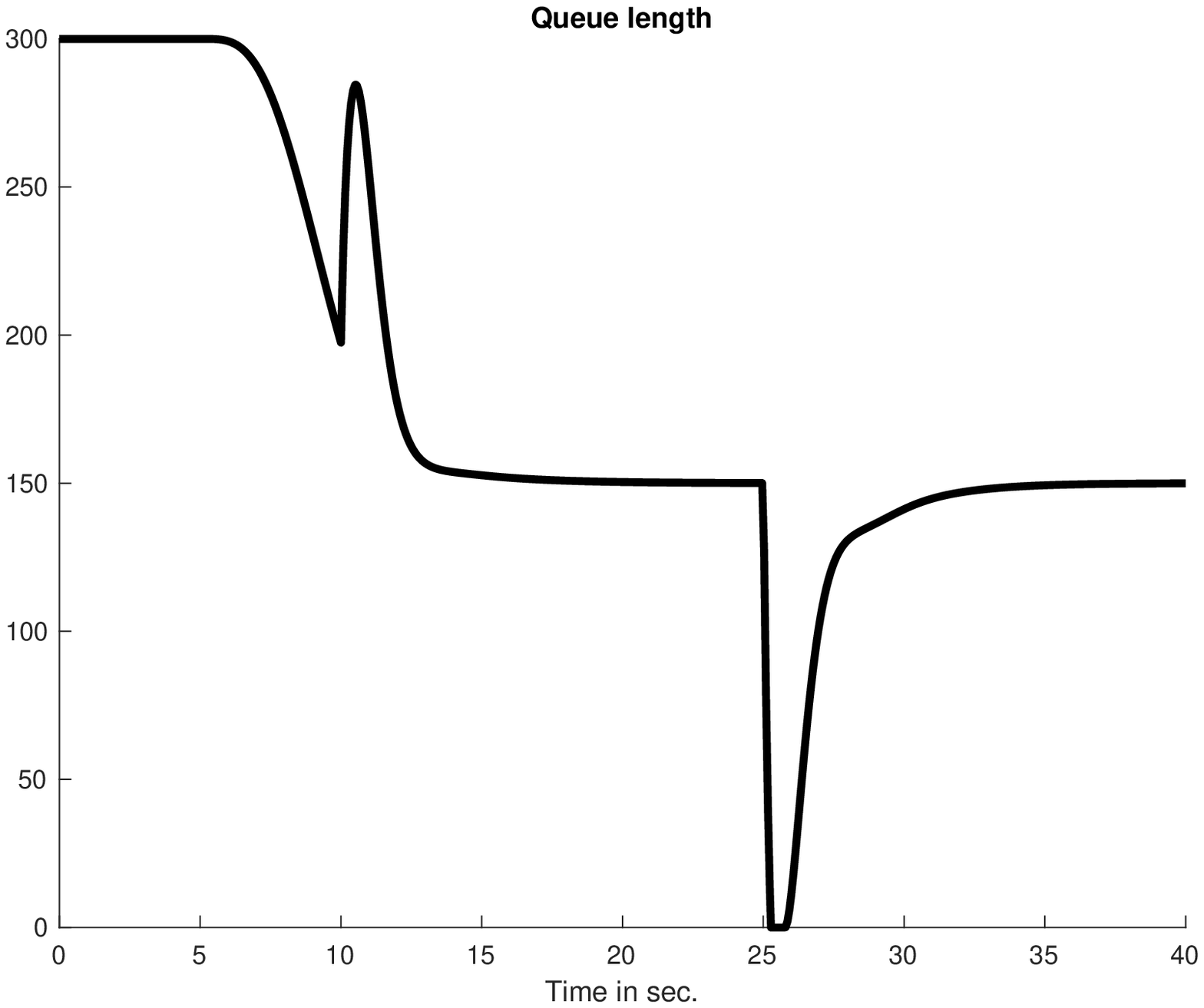,width=0.33\textwidth}}
\subfigure[\footnotesize Number of connections]
{\epsfig{figure=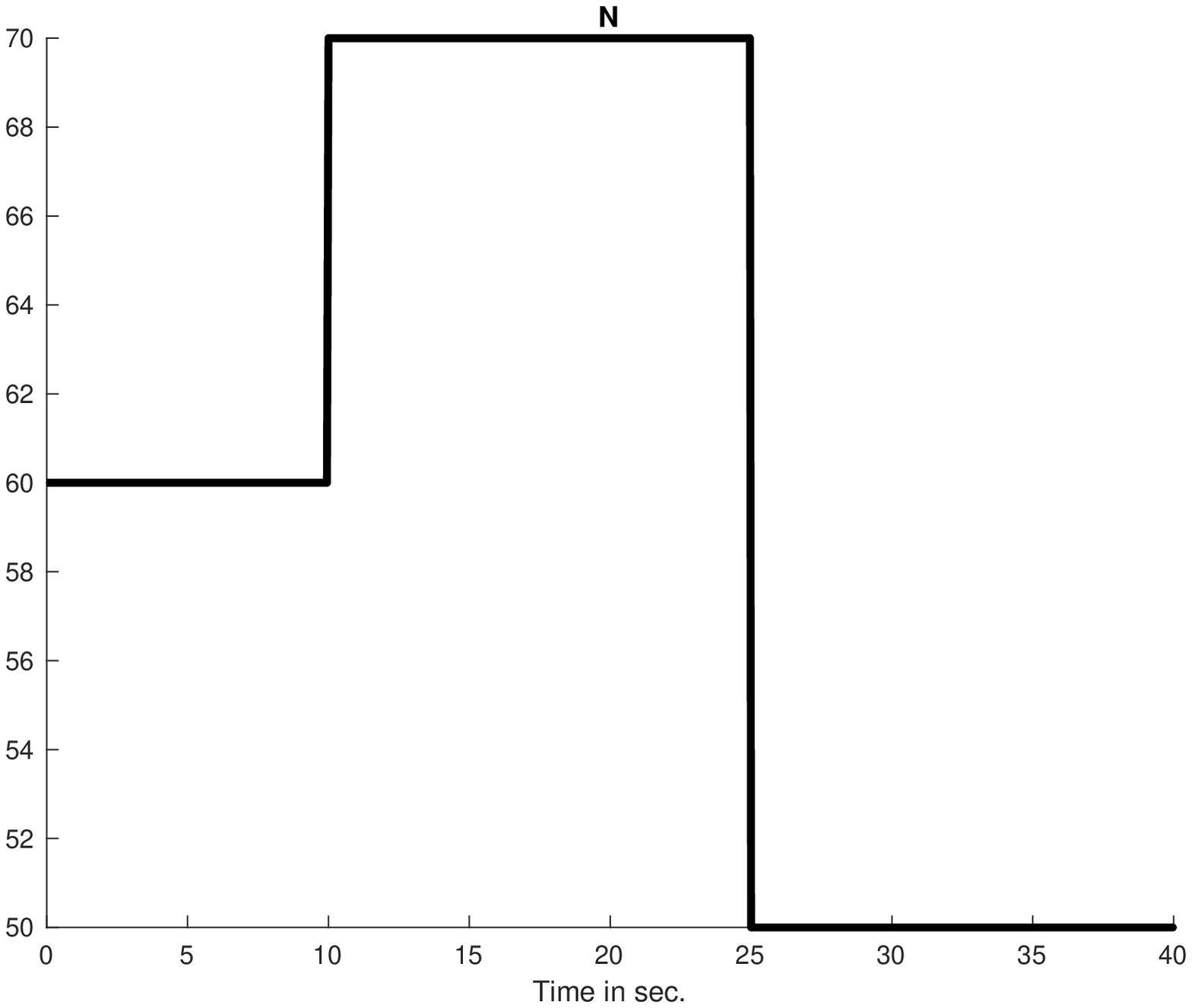,width=0.33\textwidth}}
\caption{Scenario 1 -- iPWD}\label{S24}
\end{figure*}

\begin{figure*}[!ht]
\centering%
\subfigure[\footnotesize Control (--) and nominal control (- -) ]
{\epsfig{figure=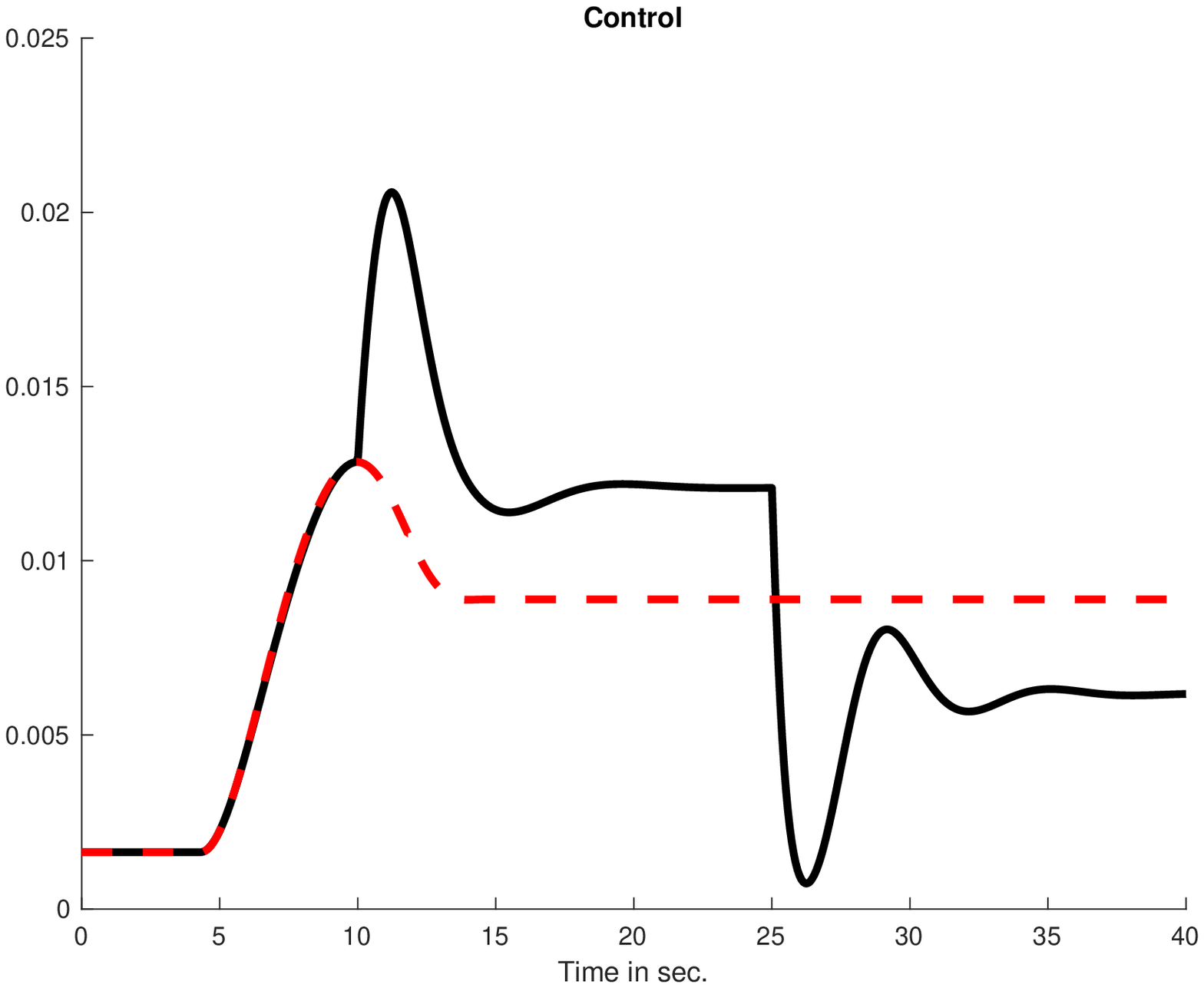,width=0.48\textwidth}}
\subfigure[\footnotesize $R$ (--) and reference trajectory (- -)]
{\epsfig{figure=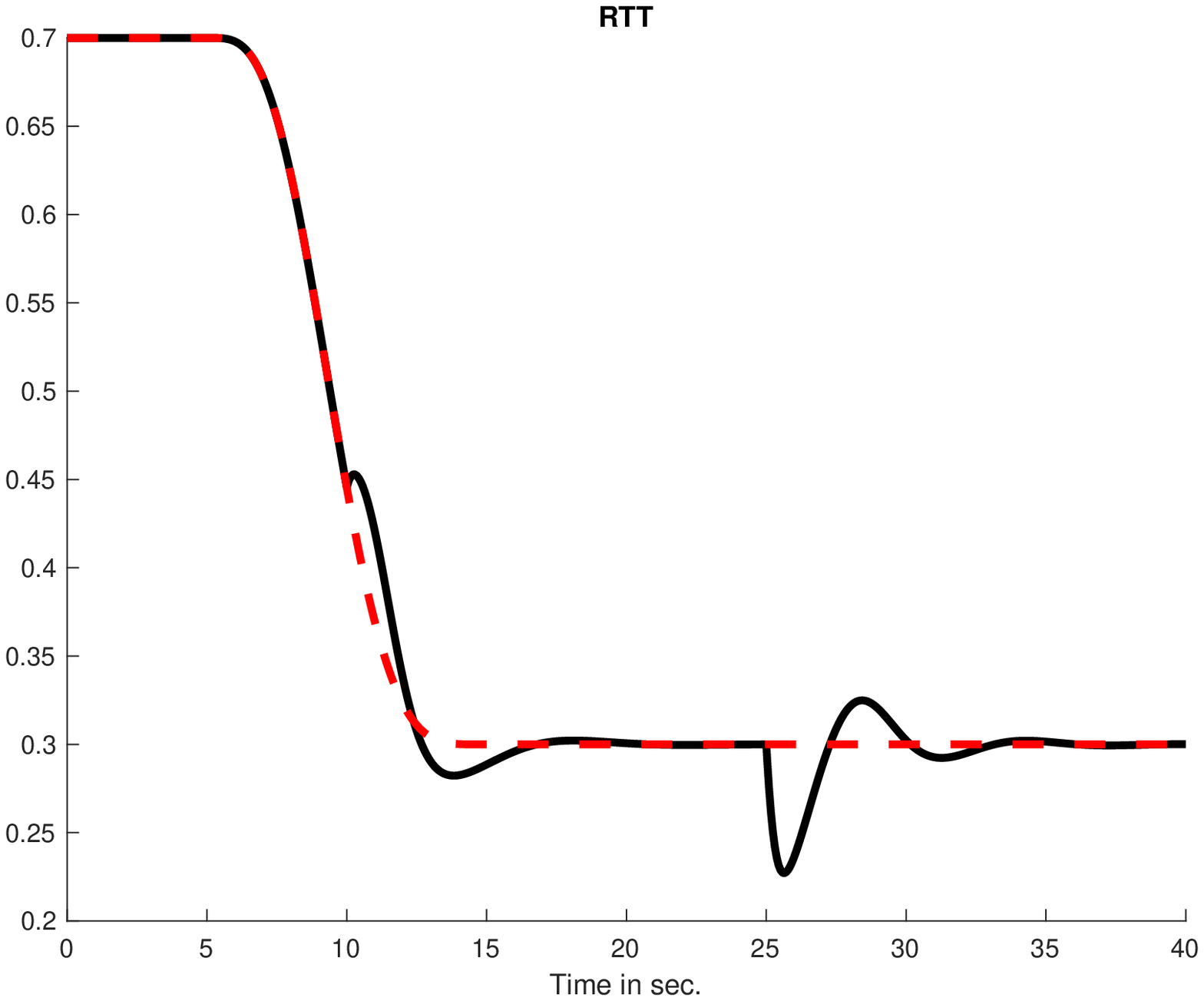,width=0.48\textwidth}}
\\
\subfigure[\footnotesize TCP Window]
{\epsfig{figure=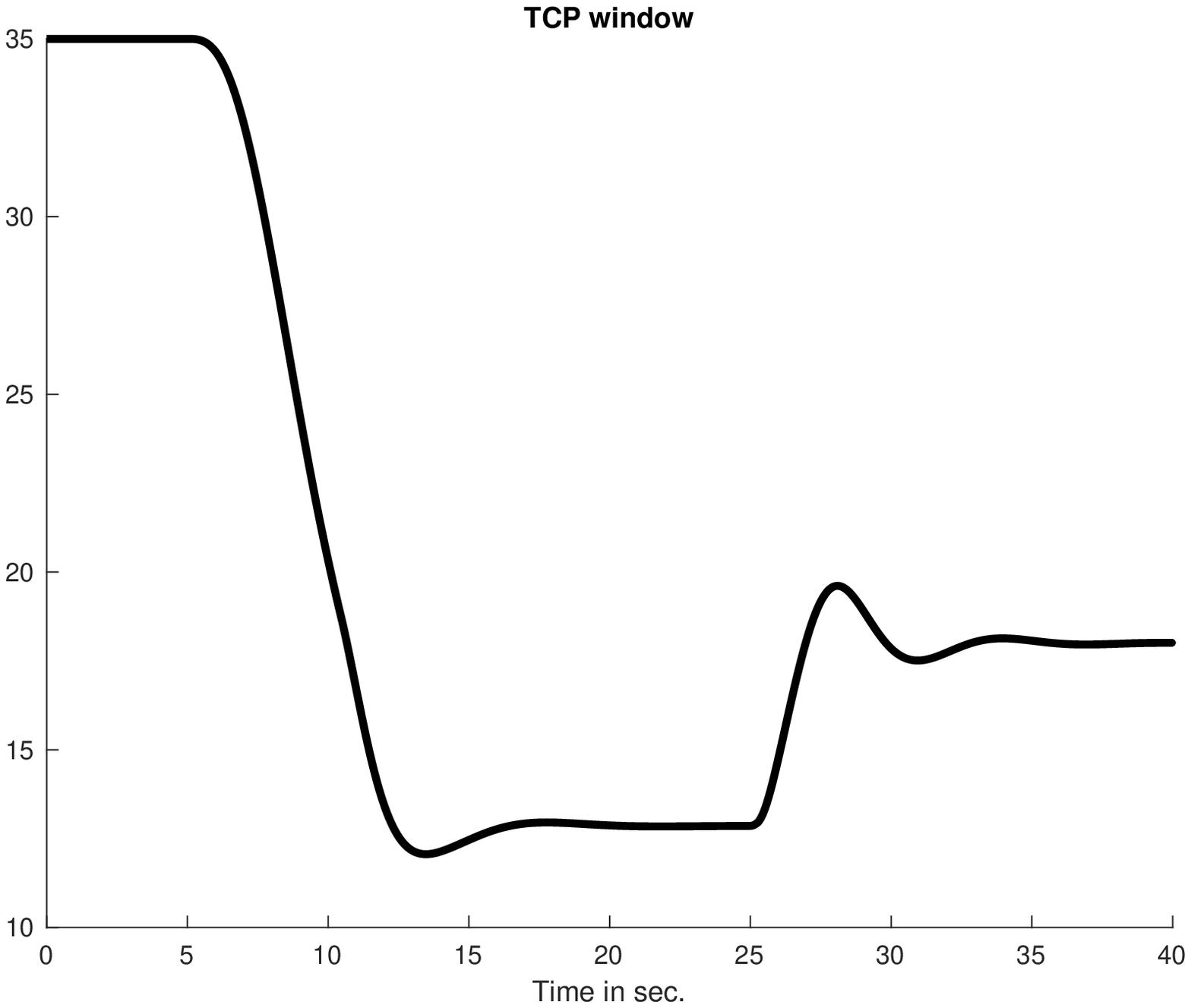,width=0.33\textwidth}}
\subfigure[\footnotesize Queue length]
{\epsfig{figure=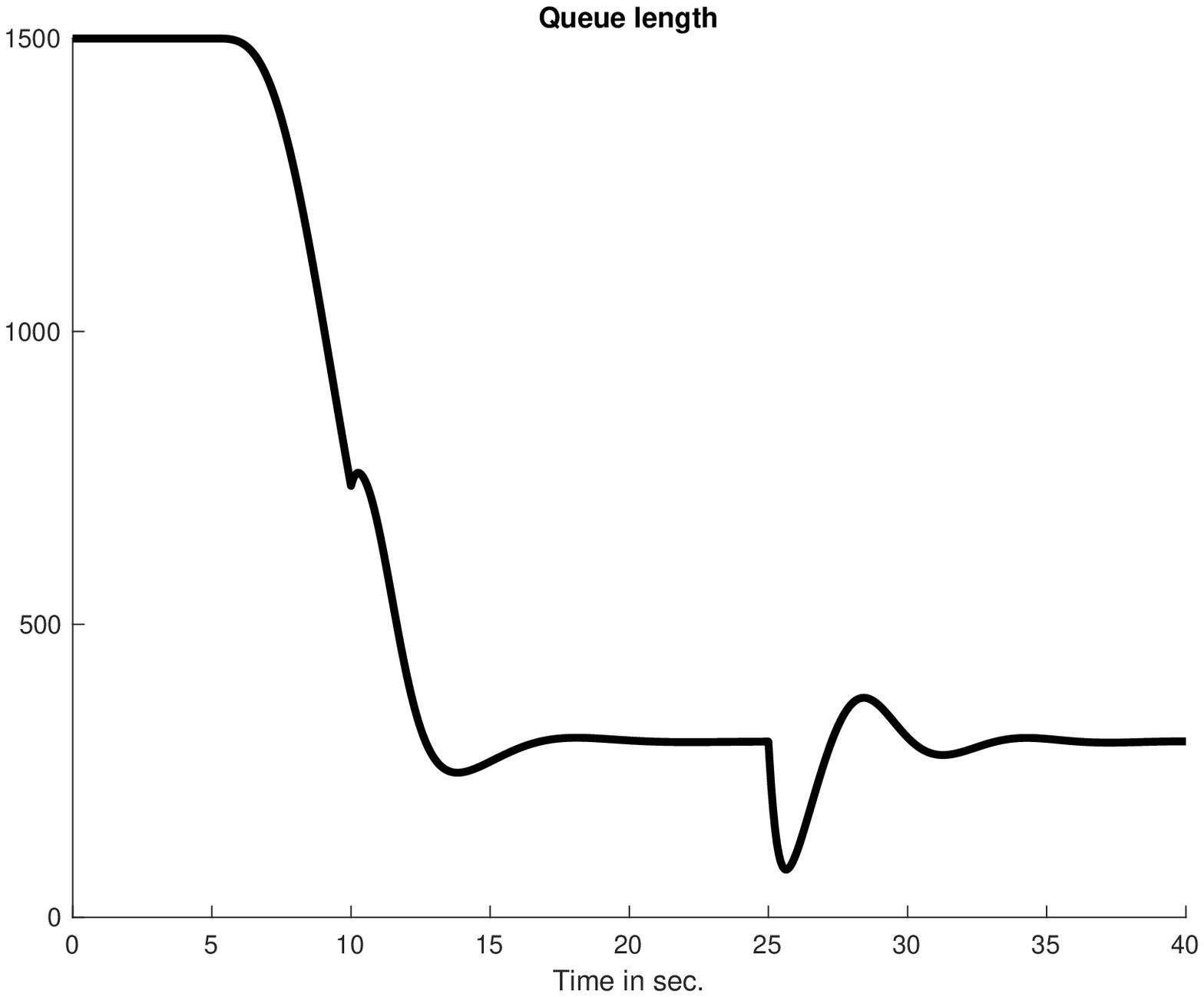,width=0.33\textwidth}}
\subfigure[\footnotesize Number of connections]
{\epsfig{figure=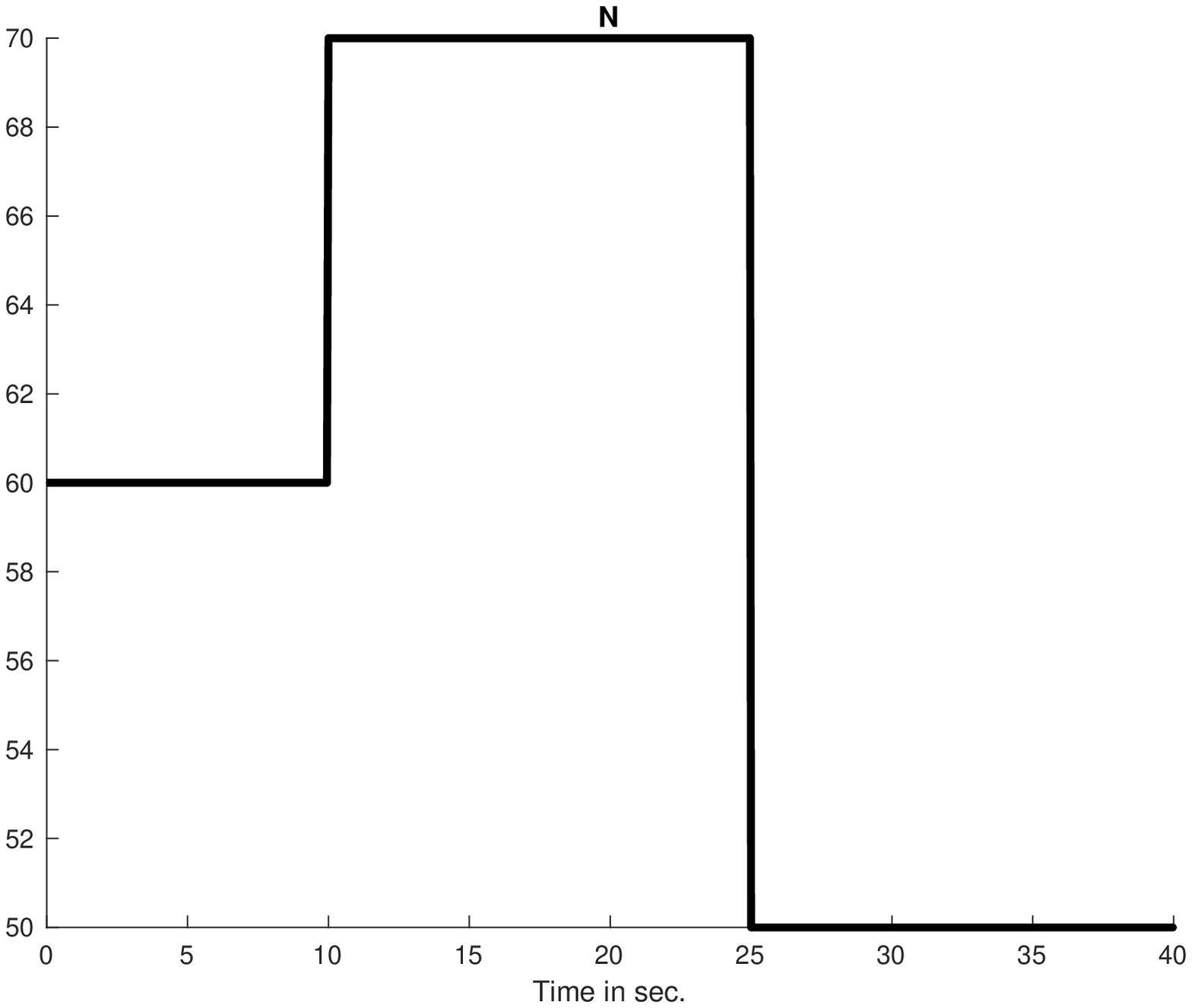,width=0.33\textwidth}}
\caption{Scenario 2 -- iP}\label{S42}
\end{figure*}
%
%
\begin{figure*}[!ht]
\centering%
\subfigure[\footnotesize Control (--) and nominal control (- -) ]
{\epsfig{figure=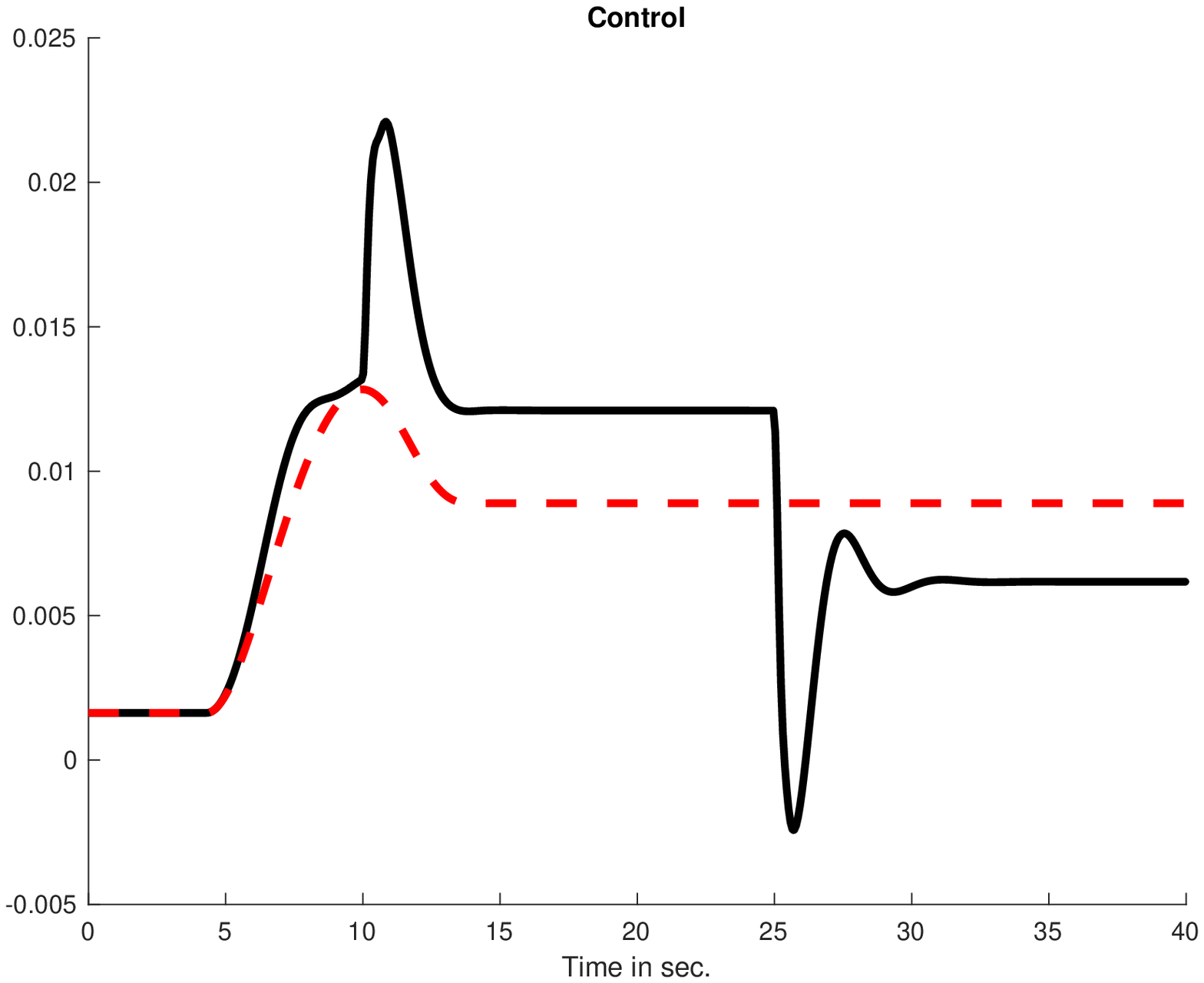,width=0.48\textwidth}}
\subfigure[\footnotesize $R$ (--) and reference trajectory (- -)]
{\epsfig{figure=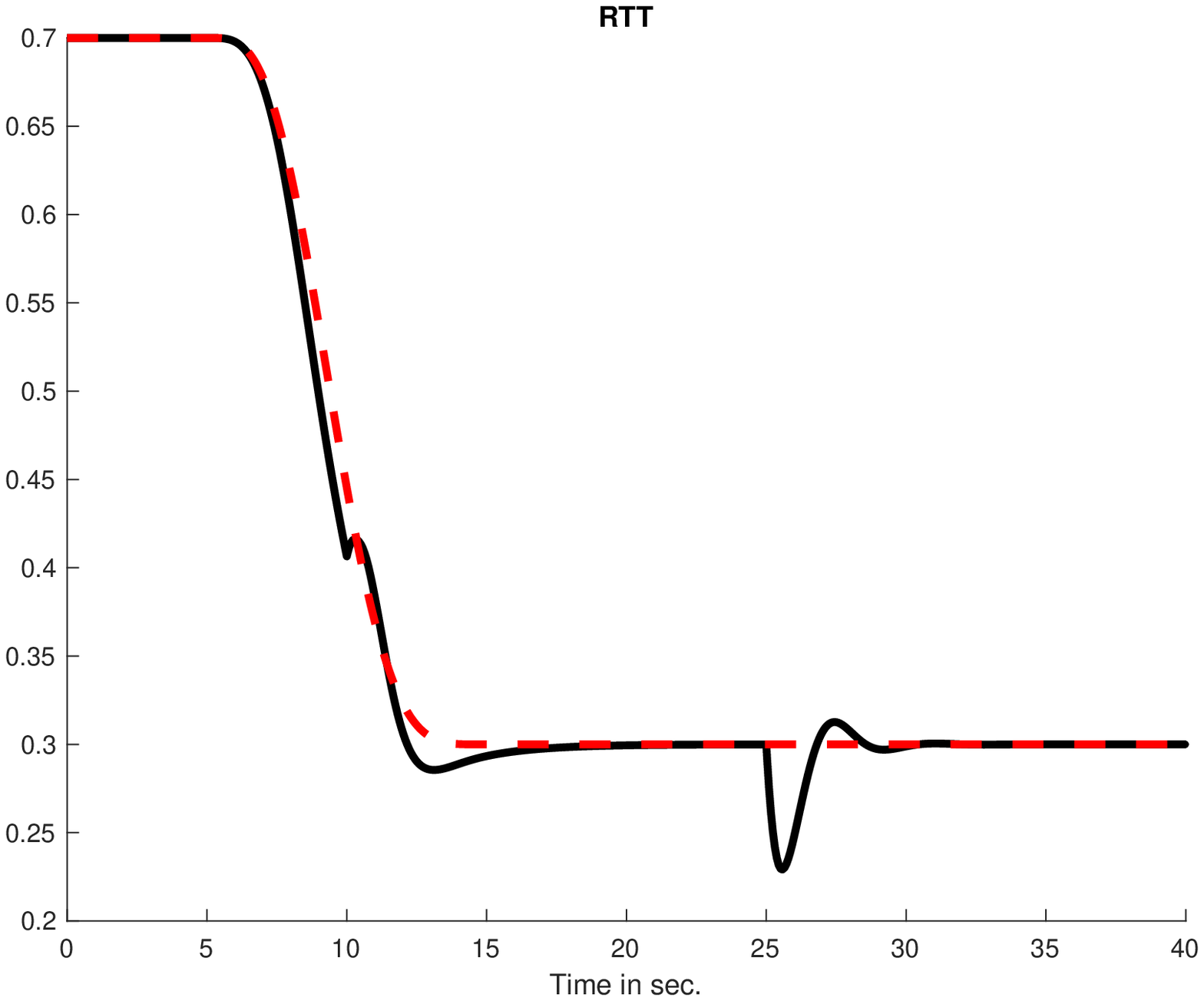,width=0.48\textwidth}}
\\
\subfigure[\footnotesize TCP Window]
{\epsfig{figure=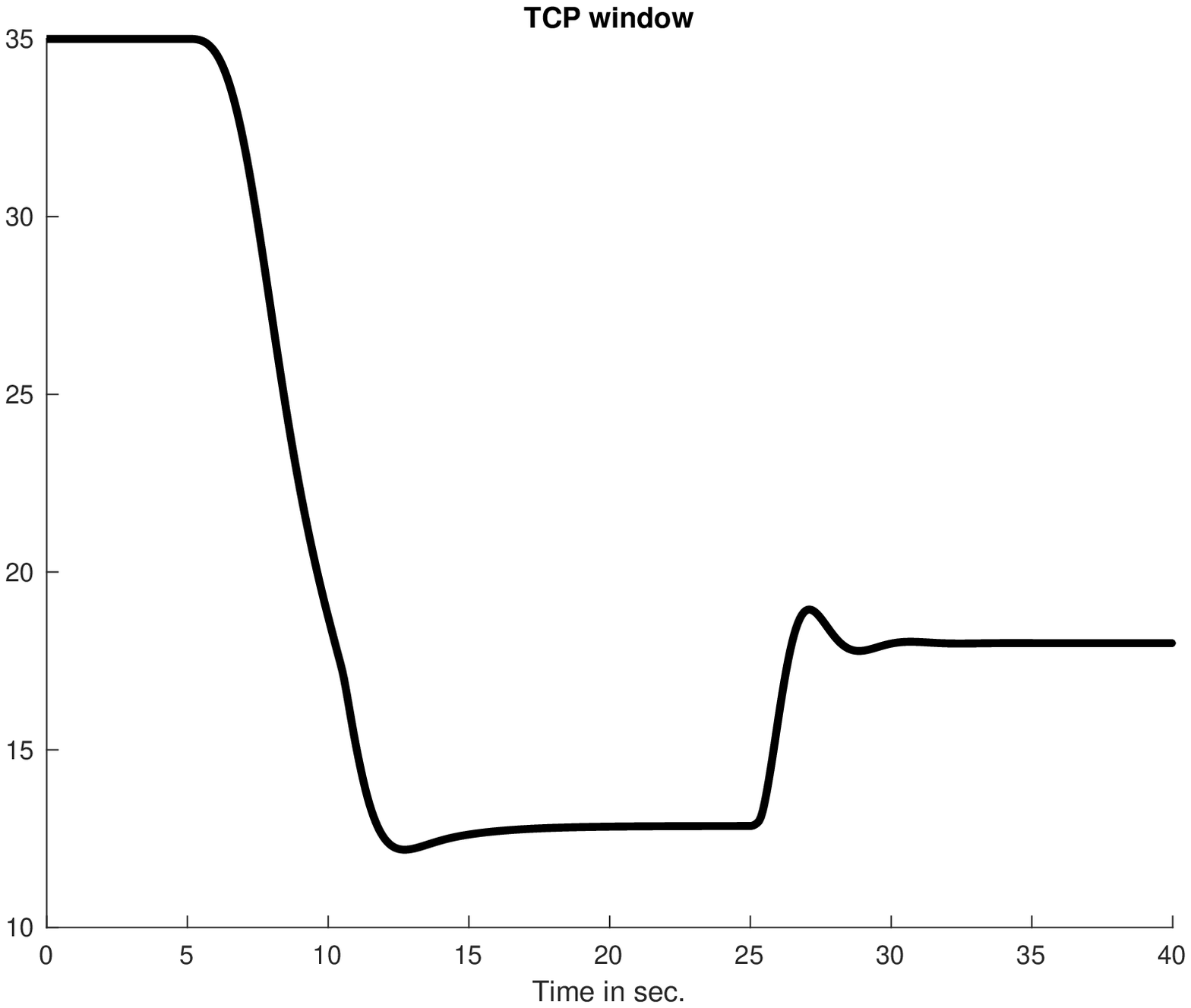,width=0.33\textwidth}}
\subfigure[\footnotesize Queue length]
{\epsfig{figure=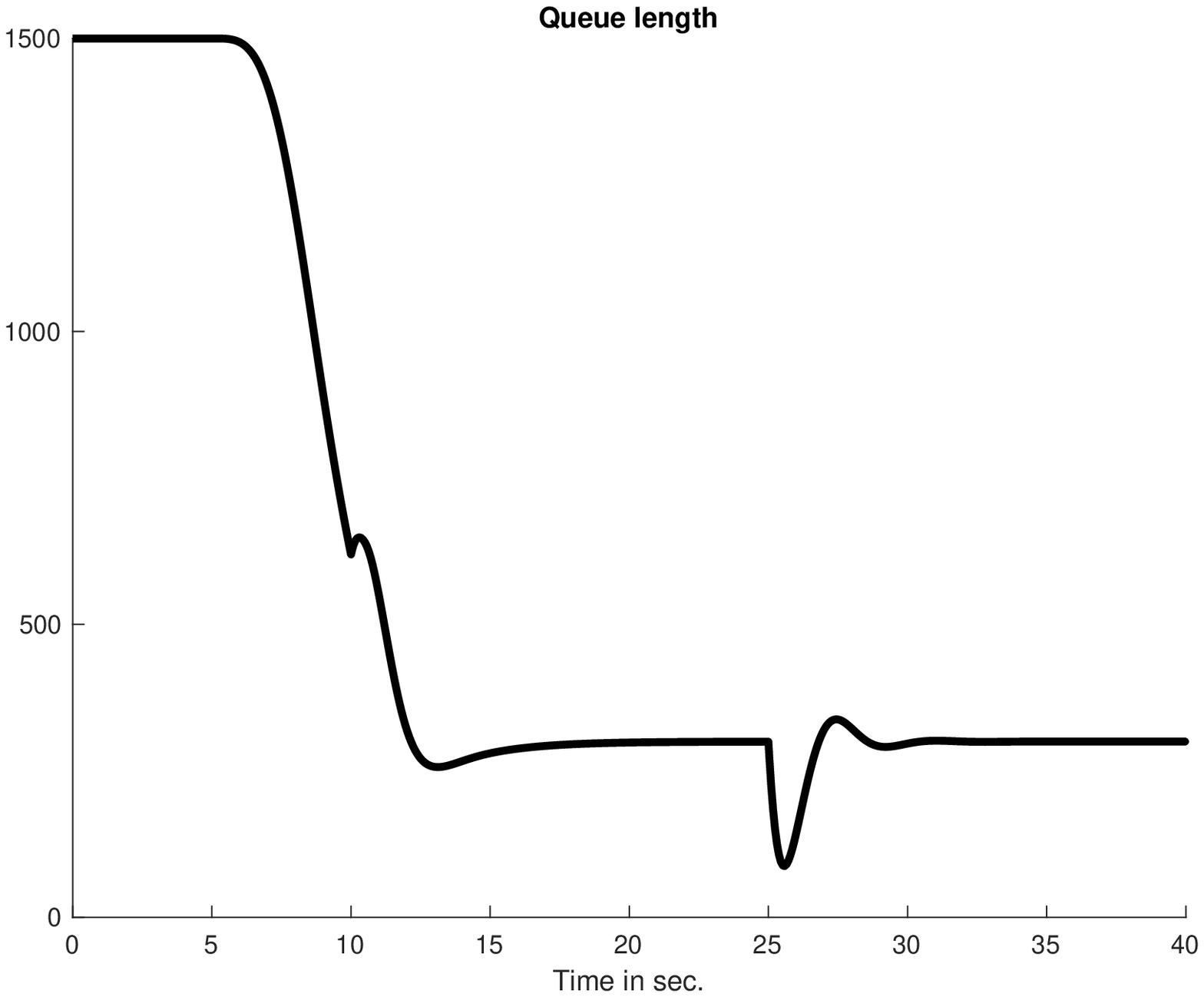,width=0.33\textwidth}}
\subfigure[\footnotesize Number of connections]
{\epsfig{figure=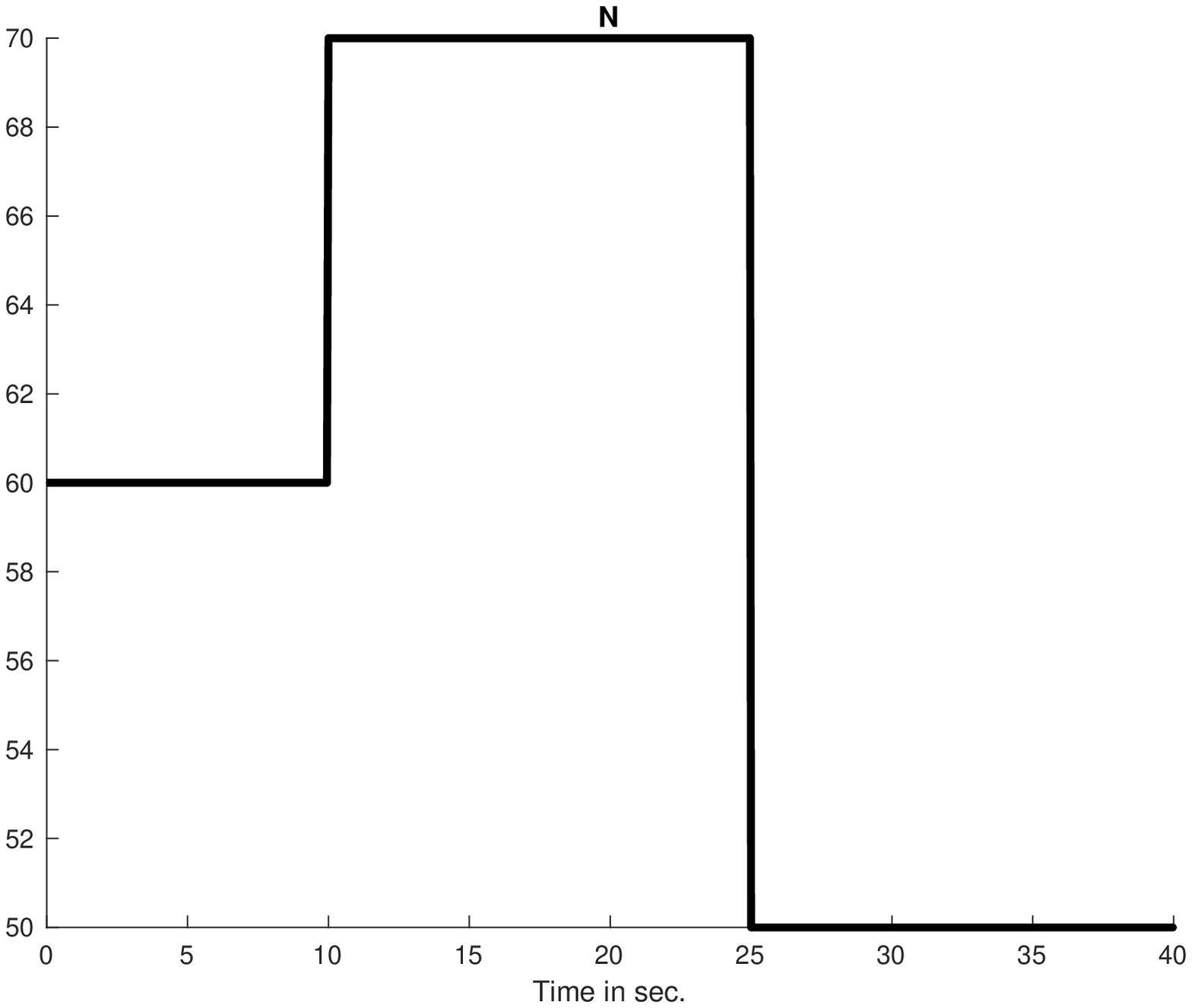,width=0.33\textwidth}}
\caption{Scenario 2 -- iPWD}\label{S44}
\end{figure*}

\clearpage





%
%
%


\end{document}